\documentclass[amsmath,amssymb,aps,pra,reprint,superscriptaddress,footnoteinbib,longbibliography]{revtex4-1}
\usepackage{amsmath,amssymb}
\usepackage[english]{babel}
\usepackage{ bbold }
\usepackage{upgreek}
\usepackage{dsfont}
\usepackage{graphicx}
\usepackage{comment}
\usepackage{braket}
\usepackage{colortbl}
\usepackage{blindtext}
\usepackage{xr-hyper}
\usepackage[hidelinks]{hyperref}
\usepackage{bm}
\allowdisplaybreaks

\begin{document}

\title{Quantum theory of non-Hermitian optical binding between nanoparticles}
\author{Henning Rudolph}
\affiliation{University of Duisburg-Essen, Faculty of Physics, Lotharstra\ss e 1, 47057 Duisburg, Germany}

\author{Uro\v{s} Deli\'c}
\affiliation{University of Vienna, Faculty of Physics, Boltzmanngasse 5, A-1090 Vienna, Austria}

\author{Klaus Hornberger}
\affiliation{University of Duisburg-Essen, Faculty of Physics, Lotharstra\ss e 1, 47057 Duisburg, Germany}

\author{Benjamin A. Stickler}
\affiliation{Ulm University, Institute for Complex Quantum Systems and Center for Integrated Quantum Science and Technology, Albert-Einstein-Allee 11, 89069 Ulm, Germany}

\begin{abstract}
Recent experiments demonstrate highly tunable nonreciprocal coupling between levitated nanoparticles due to optical binding [Rieser et al., Science {\bf 377}, 987 (2022)]. In view of recent experiments cooling nanoparticles to the quantum regime, we here develop the quantum theory of small dielectric objects interacting via the forces and torques induced by scattered tweezer photons. The interaction is fundamentally non-Hermitian  and accompanied by correlated quantum noise. We present the corresponding Markovian quantum master equation and show how to reach nonreciprocal and unidirectional coupling. Our work provides the theoretical tools for exploring and exploiting the rich quantum physics of nonreciprocally coupled nanoparticle arrays.
\end{abstract}

\maketitle

\section{Introduction}

Optically levitated nanoparticles in vacuum are a promising table-top platform for probing and exploiting quantum physics with massive objects \cite{gonzalez2021,stickler2021}. The ability to continuously monitor their dynamics, to precisely control their motion and rotation, and to let particles interact strongly in a highly tunable fashion promises a plethora of future applications in science and technology. State-of-the-art setups cool the center-of-mass motion of a single particle into its quantum ground state \cite{delic2020,magrini2021,tebbenjohanns2021,ranfagni2022,kamba2022,piotrowski2023} and rotational degrees of freedom to  millikelvin temperatures \cite{bang2020,delord2020,van2021,pontin2023}. Levitated sensors achieve force and torque sensitivities on the order of $10^{-21}$ Newton \cite{ranjit2016,hempston2017,liang2023,zhu2023} and $10^{-27}$ Newtonmeter \cite{ahn2020}. These will likely be improved further in future experiment that aim at the detection of high-frequency gravitational waves \cite{arvanitaki2013,aggarwal2022} and tests of physics beyond the standard model \cite{moore2021,carney2021b,afek2022,yin2022}. In addition, levitated nanoparticles may well allow exploring the quantum-to-classical transition at high masses \cite{bateman2014,wan2016,pino2018}, probing yet unobserved quantum interference phenomena in the rotational degrees of freedom \cite{stickler2018,ma2020,schrinski2022}, detecting nonclassical correlations in arrays of massive objects \cite{rudolph2020,brandao2021,rudolph2022,chauhan2022}, and demonstrating entanglement via Newtonian gravity \cite{bose2017,marletto2017}.

Trapping and controlling multiple objects in optical arrays is core to many future applications of levitated nanoparticles \cite{afek2022,rieser2022,brzobohaty2023,penny2023,vijayan2023}. The interference of the light scattered off one particle with the field trapping the others can give rise to strong interactions between them, commonly referred to as optical binding \cite{burns1989,karasek2006,dholakia2010,maximo2018optical}. Most experiments thus far focused on a regime where the interparticle coupling can be described as effectively conservative \cite{mohanty2004,dholakia2010,svak2021,liska2023}. However, optical binding is known to exhibit nonreciprocal behaviour \cite{obara2000,ostermann2014,sukhov2015,holzmann2015}, which seemingly violates Newton's third law. The recent study \cite{rieser2022} demonstrated full tunability between reciprocal and nonreciprocal optical binding of equally sized particles, establishing levitated nanoparticles as a viable setup for realizing non-Hermitian physics \cite{ashida2020,okuma2023}. Paradigmatic examples of effectively non-Hermitian dynamics include directional amplification \cite{wanjura2020,wang2022} and topological phase transitions \cite{shen2018,bergholtz2021,kawabata2023}, with potential applications for sensing \cite{lau2018,de2022}.

\begin{figure}[b]
	\centering
	\includegraphics[width=1\linewidth]{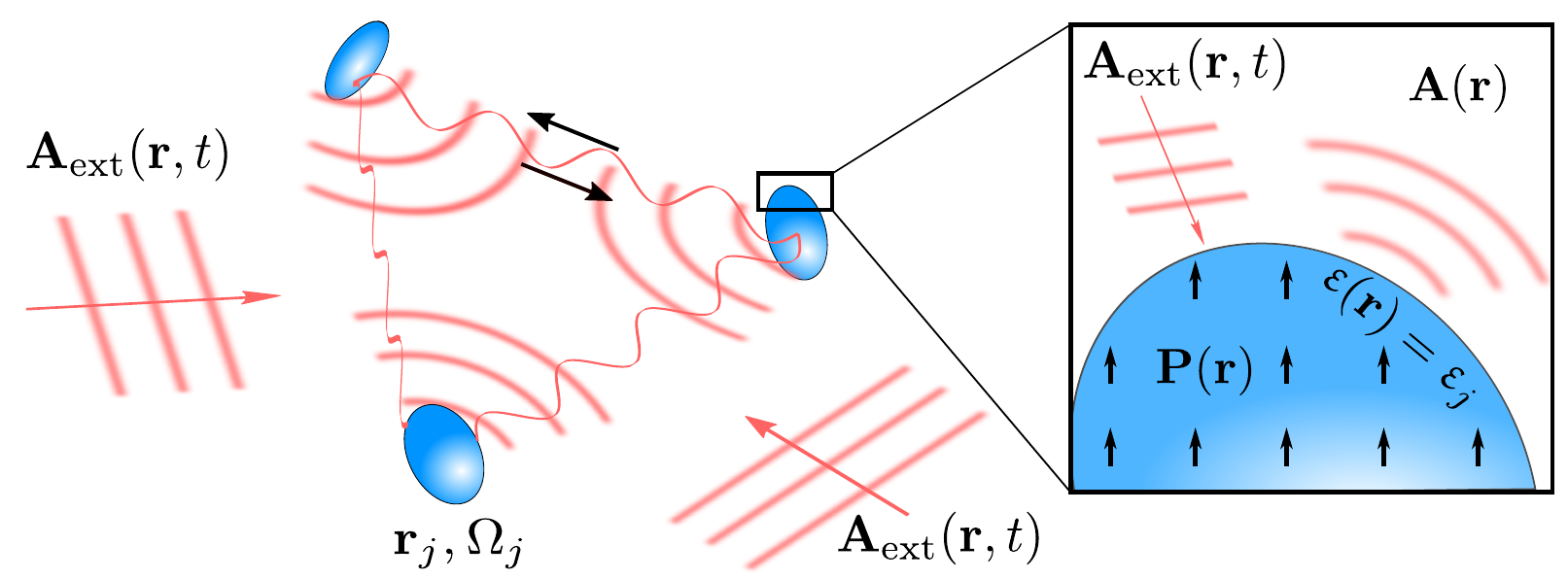}
	\caption{Several nanoparticles with center-of-mass position $\mathbf{r}_j$, orientation $\Omega_j$, and  dielectric constant $\upvarepsilon_j$ are illuminated by a laser field $\mathbf{A}_{\rm ext}(\mathbf{r},t)$. The latter induces a polarization field $\mathbf{P}(\mathbf{r})$, oscillating in phase with the laser field and leading to  dipole radiation $\mathbf{A}(\mathbf{r})$ known as Rayleigh scattering. The interference of the scattered fields with the incoming laser field gives rise to the optical-binding interaction between the particles, which may be nonreciprocal and unidirectional.}
	\label{fig:sketch}
\end{figure}

A quantum description of nonreciprocal interactions requires accounting for the fact that the two coupled partices experience correlated quantum noises \cite{kepesidis2016,metelmann2017,lau2018,zhang2019}. In the case of optical binding we show that the corresponding common bath \cite{clerk2022} is provided by the electromagnetic vacuum field surrounding the particles. Field quantization in the presence of multiple dielectrics is complicated by the fact that the total field, comprised of the incident laser and scattered radiation, must be self-consistent with the induced polarization densities, which in turn depend on the position and orientation of all particles. Solving the resulting integral equation in the Rayleigh limit of small dielectrics, we derive a Markovian quantum master equation for the coupled quantum mechanical dynamics of an arbitrary number of nanoparticles interacting via light scattering. This framework generalizes the classically observed nonreciprocal interactions \cite{arita2018,reisenbauer2023non,livska2023observations} and predicts unique signatures of quantum optical binding in terms of correlated quantum noises. The accompanying letter \cite{prl} shows how quantum optical binding can be probed and tuned in state-of-the-art experiments, paving the way for exploring and exploiting non-Hermitian quantum physics and topologically nontrivial phases in large nanoparticle arrays. In this article, we provide the theoretical framework of the quantum dynamics of optically-interacting nanoparticle arrays and discuss the prospects of generating unidirectionally coupled nanoparticles chains.

The remainder of the article is structured as follows: Section ~\ref{classicalbinding} briefly reviews the derivation of classical optical binding between nanoparticles. Then, Sec.~\ref{lightmatterinteraction} derives the quantized interaction between an arbitrary number of small dielectrics illuminated by multiple lasers and their electromagnetic environment. In Sec.~\ref{traceout}, we obtain the optical-binding quantum master equation for the particle motion by tracing out the electromagnetic vacuum. This equation is simplified in Sec.~\ref{deeplytrapped} for an array of deeply trapped nanoparticles to study the effects of correlated quantum noise and the prospects for unidirectional transport.  We discuss possible generalizations of our work in Sec.~\ref{conclusion}, and provide the quantum Langevin equations of optical binding as well as  technical details in the Appendices.

\section{Classical optical binding}\label{classicalbinding}
Before delving into the quantum description, let us briefly review the classical theory of optical binding between polarizable particles to get a better understanding of the classical dynamics we aim to reformulate in a fully quantized manner. Consider $N$  point particles at positions $\mathbf{r}_j$ with real and scalar polarizability $\alpha$. Let them be illuminated by the optical electric field $ \text{Re}\,\mathbf{E}_{\rm L}(\mathbf{r})e^{-i\omega_{\rm L}t}$, with $\omega_{\rm L}$ the laser frequency and $\mathbf{E}_{\rm L}(\mathbf{r})$ the (complex-valued) laser field profile. This field induces the dipole moment $\mathbf{p}_{0,j}(t) = \alpha \text{Re}[{\bf E}_{\rm L}({\bf r}_j) e^{-i\omega_{\rm L}t}]$ in the $j$-th particle, leading to the (complex) scattering field
\begin{align}\label{scatteringfields}
    \mathbf{E}_j(\mathbf{r}) = \frac{\alpha }{\varepsilon_0} {\sf G}(\mathbf{r}-\mathbf{r}_j){\bf E}_{\rm L}({\bf r}_j).
\end{align}
Here, we define the full electromagnetic dipole Green tensor \cite{buhmann2013}
\begin{align}\label{greentensor}
    {\sf G}(\mathbf{r}) = [\nabla\otimes\nabla + k_{\rm L}^2\mathbb{1}]\frac{e^{ik_{\rm L}r}}{4\pi r},
\end{align}
with $r=|\mathbf{r}|$. It is a Green function of the vector Helmholtz equation
\begin{equation}
\nabla \times [\nabla \times {\sf G}({\bf r})] - k^2_{\rm L} {\sf G}({\bf r}) = k^2_{\rm L} \delta({\bf r}) \mathbb{1}.
\end{equation}
For $r\neq 0$, it takes the form
\begin{align}
    {\sf G}(\mathbf{r}) = \frac{e^{ik_{\rm L}r}}{4\pi}\left( \frac{3\mathbf{r}\otimes\mathbf{r}-r^2\mathbb{1}}{r^5}(1-ik_{\rm L} r) + k_{\rm L}^2 \frac{r^2\mathbb{1}-\mathbf{r}\otimes\mathbf{r}}{r^3} \right).
\end{align}

The scattering fields of all other particles add to the laser field at the $j$-th particle position. It follows that the total induced dipole moments read approximately (neglecting multiscattering between particles) \cite{dholakia2010}
\begin{align}\label{dipolemoment}
    \mathbf{p}_j = \alpha \mathbf{E}_{\rm L}(\mathbf{r}_j) + \frac{\alpha^2}{\varepsilon_0}\sum_{\substack{j'=1\\ j'\neq j}}^N {\sf G}(\mathbf{r}_j-\mathbf{r}_{j'})\mathbf{E}_{\rm L}(\mathbf{r}_{j'}).
\end{align}
The force on particle $j$ is obtained by calculating the negative gradient of the potential of the induced dipole in the local field containing both the laser and all scattered fields
\begin{align}
    \mathbf{F}_j =  \frac{1}{2} \text{Re}\nabla\Big[\mathbf{p}_j^* \cdot\Big( \mathbf{E}_{\rm L}(\mathbf{r}) + \sum_{\substack{j'=1\\ j'\neq j}}^N \mathbf{E}_{j'}(\mathbf{r})\Big)  \Big]_{\mathbf{r}=\mathbf{r}_j},
\end{align}
where the average over one optical cycle has been taken. Inserting the dipole moment \eqref{dipolemoment} into the force and keeping only terms linear and quadratic in the particle polarizability finally leads to \cite{dholakia2010}
\begin{align}\label{classicalforces}
    \mathbf{F}_j =&\, \nabla_j \frac{\alpha}{4}|\mathbf{E}_{\rm L}(\mathbf{r}_j)|\nonumber\\ &+ \nabla_j\frac{\alpha^2}{2\varepsilon_0} \text{Re}\sum_{\substack{j'=1\\ j'\neq j}}^N \mathbf{E}_{\rm L}^*(\mathbf{r}_{j})\cdot{\sf G}(\mathbf{r}_j-\mathbf{r}_{j'})\mathbf{E}_{\rm L}(\mathbf{r}_{j'}).
\end{align}
Here, the first term describes the conservative force due to the optical potential of the $j$-th particle in the laser field, while the second term is the nonconservative optical binding force, as probed experimentally in \cite{rieser2022,reisenbauer2023non,liska2023}. Note that a full description of all optical forces would have to include the nonconservative radiation pressure force of the laser on each particle, which is on the same order of magnitude as the optical binding force \cite{rudolph2021}.

The optical binding forces \eqref{classicalforces} are not quantized straightforwardly, since they dot not share a joint potential energy \cite{rieser2022}. In the next two sections we solve this problem by first quantizing and then tracing out the total electromagnetic field.

\section{Light-matter interaction}\label{lightmatterinteraction}

\subsection{Lagrange function}\label{applagrange}
To facilitate consistent canonical quantization of several particles interacting via the electrodynamic field, we first consider the combined classical dynamics of extended dielectric matter with relative permittivity tensor $\upvarepsilon({\bf r})$ and of the electromagnetic field. The physical, i.e. real-valued, polarization field $\mathbf{P}(\mathbf{r})$ determines the density of bound charges $\rho_{\rm P}(\mathbf{r})=-\nabla \cdot \mathbf{P}(\mathbf{r})$ in the dielectric. These charges give rise to a longitudinal electric field as characterized by the {\it electrostatic} dipole Green tensor, obtainable from Eq.~\eqref{greentensor} by setting $k_{\rm L}=0$
\begin{align}\label{staticgreen}
    {\sf G}_0(\mathbf{r}) = \nabla\otimes\nabla \frac{1}{4\pi r}.
\end{align}
For $r\neq 0$ it reads
\begin{align}
    {\sf G}_0(\mathbf{r}) = \frac{3\mathbf{r}\otimes\mathbf{r}-r^2 \mathbb{1}}{4\pi r^5}.
\end{align}
In the absence of free charges, the total electric field ${\bf E}({\bf r})$ is the sum of the resulting instantaneous dipole field and of the transverse electric field, for instance due to a laser. Denoting by $\mathbf{A}_{\rm tot}(\mathbf{r})$ the vector potential in Coulomb gauge, $\nabla\cdot\mathbf{A}_{\rm tot}(\mathbf{r}) = 0$, the electric field reads
\begin{align}\label{totalfield}
    \mathbf{E}(\mathbf{r}) = -\partial_t \mathbf{A}_{\rm tot}(\mathbf{r}) + \frac{1}{\varepsilon_0} \int d^3 \mathbf{r}' {\sf G}_0(\mathbf{r}-\mathbf{r}')\mathbf{P}(\mathbf{r}').
\end{align}
Note that an integration over the Green tensor \eqref{greentensor} or \eqref{staticgreen} has to be understood as writing the gradients in front of the integral. Otherwise, the Green tensor would have to be regularized in terms of a principal value distribution \cite{van1961,yaghjian1980}.

For linear dielectrics, the internal polarization field is related to the total electric field through the constitutive relation
\begin{align}\label{eq:polfield}
    \mathbf{P}(\mathbf{r})=\varepsilon_0 [\upvarepsilon(\mathbf{r}) - \mathbb{1}]\mathbf{E}(\mathbf{r}).
\end{align}
In the following, we assume the dielectric tensor to be dispersion-free and real-valued, as applicable for light scattering off low-absorption media. Inserting $\mathbf{P}(\mathbf{r})$ into Eq.~\eqref{totalfield}, yields an integral equation for the total electric field, which is solved by
\begin{align}
    \mathbf{E}(\mathbf{r}) = -\int d^3 \mathbf{r}' {\sf L}(\mathbf{r},\mathbf{r}')\partial_t \mathbf{A}_{\rm tot}(\mathbf{r}'),
\end{align}
where the tensor-valued kernel fulfills
\begin{align}\label{KTildeinteq}
    {\sf L}(\mathbf{r},\mathbf{r}') &= \delta(\mathbf{r}-\mathbf{r}')\mathbb{1}\nonumber\\ +&\int d^3 \mathbf{s}\, {\sf G}_0(\mathbf{r}-\mathbf{s})[\upvarepsilon(\mathbf{s})-\mathbb{1}] {\sf L}(\mathbf{s},\mathbf{r}').
\end{align}
The uniqueness of its solution is guaranteed by Maxwell's equations if localized dielectrics are considered and if natural boundary conditions are assumed. The kernel ${\sf L}(\mathbf{r},\mathbf{r}')$ satisfies the symmetry relation $[\upvarepsilon(\mathbf{r})-\mathbb{1}]{\sf L}(\mathbf{r},\mathbf{r}') = {\sf L}^T(\mathbf{r}',\mathbf{r})[\upvarepsilon(\mathbf{r}')-\mathbb{1}]$.

From Eq.~\eqref{eq:polfield} the polarization field follows as
\begin{align}\label{polfield}
    \mathbf{P}(\mathbf{r}) = -\varepsilon_0 \int d^3 \mathbf{r}' {\sf K}(\mathbf{r},\mathbf{r}')\partial_t \mathbf{A}_{\rm tot}(\mathbf{r}'),
\end{align}
with the integral kernel ${\sf K}(\mathbf{r},\mathbf{r}') = [\upvarepsilon(\mathbf{r})-\mathbb{1}]{\sf L}(\mathbf{r},\mathbf{r}')$. It describes the induced electrostatic interaction between different volume elements and relates the transverse part of the field to the polarization field. The kernel vanishes outside the dielectrics, where $\upvarepsilon({\bf r}) = 1$, it is symmetric ${\sf K}(\mathbf{r},\mathbf{r}') = {\sf K}^T (\mathbf{r}',\mathbf{r})$, and fulfills the integral equation
\begin{align}\label{Kinteq}
    {\sf K}(\mathbf{r},\mathbf{r}') = &[\upvarepsilon(\mathbf{r})-\mathbb{1}]\delta(\mathbf{r}-\mathbf{r}')\nonumber\\ + &[\upvarepsilon(\mathbf{r})-\mathbb{1}] \int d^3 \mathbf{s}\, {\sf G}_0(\mathbf{r}-\mathbf{s}) {\sf K}(\mathbf{s},\mathbf{r}').
\end{align}

Now we take $\upvarepsilon({\bf r})$ to describe $N$ nonintersecting, rigid dielectric particles of arbitrary shape and size (see Fig~\ref{fig:sketch}). Denoting the center-of-mass position of the $j$-th particle by $\mathbf{r}_j$ and its orientation by $\Omega_j$, given e.g. by Euler angles $\Omega_j = (\alpha_j,\beta_j,\gamma_j)$ in the $z$-$y'$-$z''$-convention, the dielectric tensor can be written as
\begin{align}
    \upvarepsilon(\mathbf{r}) &= \mathbb{1} + \nonumber\\ &\sum_{j=1}^N {\sf R} (\Omega_j)\Big[\upvarepsilon_j[{\sf R}^T(\Omega_j)(\mathbf{r}-\mathbf{r}_j)] - \mathbb{1} \Big]{\sf R}^T(\Omega_j).
\end{align}
Here, the rotation tensors ${\sf R}(\Omega_j)$ transform from the reference orientation of the $j$-th particle to their principal axes frames so that the dielectric tensors  $\upvarepsilon_j(\mathbf{r})$ describe the individual particles in the body fixed frame. The integral kernels ${\sf K}(\mathbf{r},\mathbf{r}')$ and ${\sf L}(\mathbf{r},\mathbf{r}')$ thus depend on the center-of-mass positions and the orientation of all particles.

The total force $\mathbf{F}_j$ and the total torque $\mathbf{N}_j$ acting on the $j$-th particle follow from integrating the Lorentz force density over the particle volume ${\cal V}_j = {\cal V}_j({\bf r}_j,\Omega_j)$ (see Appendix \ref{appforce}), 
\begin{subequations}\label{forcesandtorques}
\begin{align}\label{force}
    \mathbf{F}_j = \int_{\mathcal{V}_j} d^3 \mathbf{r}\, \nabla'[\mathbf{P}(\mathbf{r})\cdot \mathbf{E}(\mathbf{r}')]_{\mathbf{r}' = \mathbf{r}}
\end{align}
and
\begin{align}\label{torque}
    \mathbf{N}_j =& \int_{\mathcal{V}_j} d^3 \mathbf{r}\, \mathbf{P}(\mathbf{r})\times \mathbf{E}(\mathbf{r})\nonumber\\ &+ (\mathbf{r}-\mathbf{r}_j) \times \nabla' [\mathbf{P}(\mathbf{r})\cdot \mathbf{E}(\mathbf{r}')]_{\mathbf{r}' = \mathbf{r}}.
\end{align}
Note that the gradients in Eqs.~(\ref{force}) and (\ref{torque}) act only on the electric field, reflecting that the potential energy of a polarized volume element $d^3 \mathbf{r}$ with constant dipole moment $\mathbf{P}(\mathbf{r}) d^3\mathbf{r}$ is $-\mathbf{P}(\mathbf{r})\cdot \mathbf{E}(\mathbf{r}) d^3\mathbf{r}$. The first term in Eq.~(\ref{torque}) is the intrinsic torque on each volume element, describing the precession of the dipole density in the electric field, while the second term is caused by the force density acting on the particle, inducing an orbital torque around the particle center of mass.

These equations are complemented by the wave equation of the transverse vector potential sourced by the transverse part of the polarization current density $\partial_t\mathbf{P}(\mathbf{r})$ \cite{jackson1999},
\begin{align}\label{transversecurrent}
\left    (\frac{1}{c^2}\partial_t^2 - \Delta \right )\mathbf{A}_{\rm tot}(\mathbf{r}) = \mu_0 \partial_t \mathbf{P}_\perp(\mathbf{r}).
\end{align}
\end{subequations}

The equations of motion \eqref{forcesandtorques} can be formulated as Euler-Lagrange equations with Lagrangian
\begin{align}\label{lagrangestart}
    &L_{\rm tot} = L_{\rm m} + L_{\rm em}^{\rm tot} + L_{\rm int}.
\end{align}
Here, $L_{\rm m}$ denotes the free mechanical Lagrangian of all particles, including their translational and rotational kinetic energies as well as possible external potentials. The free electromagnetic Lagrangian of the transverse electromagnetic field \cite{schubert1993} reads
\begin{align}\label{freefield}
    L_{\rm em}^{\rm tot} = \int d^3\mathbf{r}\left(\frac{\varepsilon_0}{2}\left[ \partial_t \mathbf{A}_{\rm tot}(\mathbf{r}) \right]^2 - \frac{1}{2\mu_0} \left[ \nabla\times\mathbf{A}_{\rm tot}(\mathbf{r}) \right]^2\right).
\end{align}
Finally, the light-matter interaction is accounted for by
\begin{align}\label{lagrangeint}
    L_{\rm int} = \frac{\varepsilon_0}{2} \int d^3\mathbf{r} d^3\mathbf{r}' \left[ \partial_t \mathbf{A}_{\rm tot}(\mathbf{r}) \right]\cdot {\sf K}(\mathbf{r},\mathbf{r}') \left[ \partial_t \mathbf{A}_{\rm tot}(\mathbf{r}') \right].
\end{align}
Note that this can be understood as the energy of the polarization field in the transverse electric field characterized by the energy density $\mathbf{P}(\mathbf{r})\cdot \partial_t \mathbf{A}_{\rm tot}(\mathbf{r})/2$.

The above derivation assumes rigid dielectrics, which move and revolve on a timescale slow compared to the light propagation through them. Extending this treatment to include other mechanical degrees of freedom, such as elasto-mechanic deformations of the bodies, is conceptually straightforward. The resulting Lagrangian takes the form Eq.~\eqref{lagrangestart}, but with the integral kernel depending also on additional generalized coordinates.

The fact that the interaction Lagrangian \eqref{lagrangeint} describes the full light-matter coupling will prove crucial for quantizing the theory in the small particle limit. This coupling term differs from that typically used for levitated particles \cite{romero2011,stickler2016b,gonzalez2019}, even though the difference becomes relevant only when considering more than a single particle.

\subsection{Euler-Lagrange equations}
We now confirm that the Lagrangian (\ref{lagrangestart}) yields the equations of motion \eqref{forcesandtorques} for rigid objects. This means that the forces \eqref{force} must be given by 
\begin{align}\label{eq:eulerlagrange}
    {\bf F}_j = \partial_{{\bf r}_j}L_{\rm int}.
\end{align}
The derivative on the right-hand side acts only on the integral kernel ${\sf K}(\mathbf{r},\mathbf{r}')$. Denoting its $\ell$-th cartesian component by $({\bf e}_\ell \cdot \partial_{\mathbf{r}_j})$, we first use that
\begin{align}
    ({\bf e}_\ell \cdot \partial_{\mathbf{r}_j}) [\upvarepsilon(\mathbf{r}) - \mathbb{1}] = \begin{cases}
- \nabla_\ell [\upvarepsilon(\mathbf{r}) - \mathbb{1}] & {\rm for} \quad {\bf r}\in {\cal V}_j \\
0 & {\rm elsewhere.}
    \end{cases}
\end{align}
Then, in order to evaluate $({\bf e}_\ell \cdot \partial_{\mathbf{r}_j}) {\sf L}(\mathbf{r},\mathbf{r}')$ we apply the derivative to Eq.~\eqref{KTildeinteq} and solve the latter,
\begin{align}
    ({\bf e}_\ell \cdot \partial_{\mathbf{r}_j}) {\sf L}(\mathbf{r},\mathbf{r}') = & \int d^3 \mathbf{s}' d^3 \mathbf{s} \, {\sf L}(\mathbf{r},\mathbf{s}') {\sf G}_0(\mathbf{s}' - \mathbf{s})({\bf e}_\ell\cdot\partial_{\mathbf{r}_j})\nonumber\\ 
    & [\upvarepsilon(\mathbf{s})-\mathbb{1}] {\sf L}(\mathbf{s},\mathbf{r}').
\end{align}
Integration by parts and identifying the integral kernels (\ref{KTildeinteq}) and (\ref{Kinteq}) finally leads to
\begin{align}
    {\bf e}_\ell\cdot \partial_{{\bf r}_j} L_{\rm int} = & \varepsilon_0\int d^3 \mathbf{s}' d^3 \mathbf{s} \int_{\mathcal{V}_j} d^3\mathbf{r} \partial_t \mathbf{A}_{\rm tot}(\mathbf{s}')\nonumber\\ &\cdot {\sf K}(\mathbf{s}',\mathbf{r}) \nabla_\ell {\sf L}(\mathbf{r},\mathbf{s}) \partial_t \mathbf{A}_{\rm tot}(\mathbf{s}),
\end{align}
which demonstrates Eq.~\eqref{eq:eulerlagrange}.

In order to show that the Lagrangian yields the torque \eqref{torque} we must evaluate the derivative of $L_{\rm int}$ with respect to the Euler angles $\mu_j \in \lbrace \alpha_j,\beta_j,\gamma_j \rbrace$ of the $j$-th particle. We proceed as for the center of mass to obtain
\begin{align}
    \partial_{\mu_j} {\sf L}(\mathbf{r},\mathbf{r}') = & \int d^3 \mathbf{s}' d^3 \mathbf{s}\, {\sf L}(\mathbf{r},\mathbf{s}')\nonumber\\ &{\sf G}_0(\mathbf{s}' - \mathbf{s}) \partial_{\mu_j} [\upvarepsilon(\mathbf{s})-\mathbb{1}] {\sf L}(\mathbf{s},\mathbf{r}').
\end{align}
The integration can be restricted to the $j$-th particle volume, yielding
\begin{align}\label{torquemid}
    \partial_{\mu_j} L_{\rm int} = \frac{\varepsilon_0}{2}\int_{\mathcal{V}_j} d^3 \mathbf{r}\, \mathbf{E}(\mathbf{r})\cdot \partial_{\mu_j}[ \upvarepsilon(\mathbf{r}) -\mathbb{1}]\mathbf{E}(\mathbf{r}),
\end{align}
where the derivative only acts on the dielectric tensor. Using $\partial_{\mu_j}{\sf R}(\Omega_j) = \mathbf{n}_{\mu_j} \times {\sf R}(\Omega_j)$, where $\mathbf{n}_{\mu_j}$ is the instantaneous rotation axis associated with $\mu_j$, one gets
\begin{align}\label{epsilonderivative}
    \partial_{\mu_j}[\upvarepsilon(\mathbf{r})-\mathbb{1}] =& \mathbf{n}_{\mu_j} \times [\upvarepsilon(\mathbf{r}) - \mathbb{1}] + \Big[\mathbf{n}_{\mu_j} \times [\upvarepsilon(\mathbf{r}) - \mathbb{1}]\Big]^T \nonumber\\ &+ [(\mathbf{r}-\mathbf{r}_j)\cdot( \mathbf{n}_{\mu_j} \times \nabla)][\upvarepsilon(\mathbf{r})-\mathbb{1}].
\end{align}
Inserting this into Eq.~\eqref{torquemid}, the first line in Eq.~\eqref{epsilonderivative} yields the intrinsic torque while the second line, after partial integration, yields the torque due to the force density. In total, we thus showed that
\begin{align}
     \partial_{\mu_j} L_{\rm int} = \mathbf{n}_{\mu_j}\cdot \mathbf{N}_j,
\end{align}
which is Eq.~(\ref{torque}).

Finally, also the right-hand side of the wave equation \eqref{transversecurrent} must follow from Eq.~(\ref{lagrangeint}). As $L_{\rm int}$ depends only on the time derivative of the transverse vector potential, we need the functional derivative of $L_{\rm int}$ only with respect to $\partial_t \mathbf{A}_{\rm tot}$. Taking the transversality of the vector potential into account, one obtains
\begin{align}
    \partial_t \frac{\delta L_{\rm int}}{\delta [\partial_t \mathbf{A}_{\rm tot}(\mathbf{r})]} = \varepsilon_0 \partial_t \left( \int d^3 \mathbf{r}' {\sf K}(\mathbf{r},\mathbf{r}')\partial_t\mathbf{A}_{\rm tot}(\mathbf{r}') \right)_\perp.
\end{align}
Expressing the right-hand side through the transverse part of the polarization field Eq.~\eqref{polfield}, one obtains Eq.~(\ref{transversecurrent}).

\subsection{Small particles in an external field}\label{smallellipsoids}
From now on we assume for simplicity that all arbitrarily shaped particles exhibit a uniform and isotropic permittivity tensor $\varepsilon \mathbb{1}$ and that the particles are much smaller than the distances between them. To treat the inter-particle interactions as a perturbation in the integral kernel \eqref{Kinteq}, we first define the integral kernel ${\sf K}_0(\mathbf{r},\mathbf{r}')$ for the case that all particles are arbitrarily far separated from one another. It fulfills the integral equation
\begin{align}\label{Kimpeq}
    {\sf K}_0(\mathbf{r},\mathbf{r}') = &(\varepsilon-1)\delta(\mathbf{r}-\mathbf{r}')\mathbb{1}\nonumber\\& + (\varepsilon-1) \int_{\mathcal{V}_j} d^3\mathbf{s}\, {\sf G}_0(\mathbf{r}-\mathbf{s}){\sf K}_0(\mathbf{s},\mathbf{r}')
\end{align}
for $\mathbf{r},\mathbf{r'}\in \mathcal{V}_j$ and vanishes everywhere else. The dependence on all canonical coordinates, collectively denoted by $q_j$ from now on, enters via the regions $\mathcal{V}_j$ inhabited by the particles.

We will focus on particles of ellipsoidal shape, see Fig.~\ref{fig:sketch}. All external fields are approximately constant in the particle volumes, such that the integral equation \eqref{Kimpeq} can be solved for ellipsoids as \cite{rudolph2021,hulst1981}
\begin{align}\label{unperturbed}
    {\sf K}_0(\mathbf{r},\mathbf{r}') = \begin{cases}
    \upchi_j \delta(\mathbf{r}-\mathbf{r}') &\text{ for } \mathbf{r},\mathbf{r}'\in\mathcal{V}_j \\ 0 &\text{ else},
    \end{cases}
\end{align}
with the susceptibility tensors
\begin{align}\label{eq:suscepttsensor}
    \upchi_j = \frac{\varepsilon - 1}{1+{\sf N}_j(\varepsilon - 1)},
\end{align}
involving the depolarization tensors ${\sf N}_j$, which depend only on the particle diameters along their principal axes $(\ell_{j,1},\ell_{j,2},\ell_{j,3})$. The eigenvalues of the depolarization tensors read
\begin{align}\label{depol}
    N_{j,1} = \frac{\ell_{j,1} \ell_{j,2} \ell_{j,3}}{2}\int_0^{\infty} \frac{ds}{\sqrt{(s+\ell_{j,1}^2)^3(s+\ell_{j,2}^2)(s+\ell_{j,3}^2)}}.
\end{align}
The other two eigenvalues $N_{j,2}$ and $N_{j,3}$ follow by a permutation of the second index. The tensors fulfill $\text{tr}\lbrace {\sf N_j} \rbrace = 1$ and are rotated according to the particle orientation.

We now define the inter-particle interaction kernel ${\sf K}_{\rm int}(\mathbf{r},\mathbf{r}')$ by decomposing ${\sf K}(\mathbf{r},\mathbf{r}')$ into 
\begin{align}
    {\sf K}(\mathbf{r},\mathbf{r}') = {\sf K}_0(\mathbf{r},\mathbf{r}') + {\sf K}_{\rm int}(\mathbf{r},\mathbf{r}').
\end{align}
Inserting this into Eq.~\eqref{Kinteq} and treating the interaction as a small perturbation, we get an integral equation for ${\sf K}_{\rm int}(\mathbf{r},\mathbf{r}')$. For $\mathbf{r} \in \mathcal{V}_j$ it reads
\begin{align}
    {\sf K}_{\rm int}(\mathbf{r},\mathbf{r}') = &(\varepsilon - 1)\sum_{\substack{j'=1\\ j'\neq j}}^N\Theta_{j'}(\mathbf{r}') {\sf G}_0(\mathbf{r}-\mathbf{r}')\upchi_{j'}\nonumber\\& + (\varepsilon-1) \int_{\mathcal{V}_j} d^3\mathbf{s}\, {\sf G}_0(\mathbf{r}-\mathbf{s}){\sf K}_{\rm int}(\mathbf{s},\mathbf{r}'),
\end{align}
involving the indicator functions $\Theta_j(\mathbf{r})$, which take unit value inside the $j$-th particle volume and vanish otherwise. As the particle distances are always greater than the particles, such that all dipole fields of other particles are approximately constant along the $j$-th particle volume, ${\sf K}_{\rm int}$ can be obtained by using the same steps that lead to Eq.~(\ref{unperturbed}) as
\begin{align}\label{perturbed}
    {\sf K}_{\rm int}(\mathbf{r},\mathbf{r}') = \begin{cases}
    \upchi_j {\sf G}_0(\mathbf{r}-\mathbf{r}')\upchi_{j'}&\text{ for } \mathbf{r}\in\mathcal{V}_j \text{ and } \mathbf{r}'\in \mathcal{V}_{j'} \\ &\text{ with }j\neq j', \\ 0 &\text{ else}.
    \end{cases}
\end{align}
Inserting the thus obtained integral kernel ${\sf K}(\mathbf{r},\mathbf{r}')$ into Eq.~\eqref{lagrangeint}, the light-matter interaction decomposes into an optical potential-type term due to ${\sf K}_0(\mathbf{r},\mathbf{r}')$, describing the energy of the induced dipoles within the external electromagnetic field, and the electrostatic interaction between different induced dipoles due to ${\sf K}_{\rm int}(\mathbf{r},\mathbf{r}')$.

Next, we write the electromagnetic vector potential as the superposition of a classical, transverse, externally given electromagnetic field $\mathbf{A}_{\rm ext}(\mathbf{r},t)$, describing the external laser light illuminating the particles, and a dynamical field $\mathbf{A}(
\mathbf{r})$, describing the light scattered by the particles (see Fig.~\ref{fig:sketch}),
\begin{align}\label{fieldsplit}
    \mathbf{A}_{\rm tot}(\mathbf{r}) = \mathbf{A}_{\rm ext}(\mathbf{r},t) + \mathbf{A}(\mathbf{r}).
\end{align}
The dynamical part of the field $\mathbf{A}(\mathbf{r})$ will later be quantized. Throughout the article we drop the explicit time dependence of all dynamical variables and fields such as $\mathbf{r}_j,\mathbf{A}_{\rm tot}(\mathbf{r}),\mathbf{A}(\mathbf{r})$, but keep it for externally prescribed functions, such as for $\mathbf{A}_{\rm ext}(\mathbf{r},t)$. The external vector potential fulfills the homogeneous wave equation,
\begin{align}\label{freewave}
    \left( \frac{1}{c^2}\partial_t^2 - \Delta \right) \mathbf{A}_{\rm ext}(\mathbf{r},t) = 0.
\end{align}

Choosing $\mathbf{A}_{\rm ext}(\mathbf{r},t)$ such that all relevant wavelengths of its spectral representation are much greater than the particle sizes, we can neglect all light-matter interaction terms quadratic in the scattering fields $\mathbf{A}(\mathbf{r})$ in $L_{\rm tot}$ and approximate $\mathbf{A}_{\rm tot}(\mathbf{r})$ as the laser field $\mathbf{A}_{\rm ext}(\mathbf{r},t)$ in the electrostatic interaction described by ${\sf K}_{\rm int}(\mathbf{r},\mathbf{r}')$. Additionally, since the particles move slowly compared to the speed of light, such that the latter adapts instantaneously to a new particle state, the Lagrangian (\ref{lagrangestart}) can be written as $L_{\rm tot} = L - dS/dt$. Then, the Lagrangian $L$ reads
\begin{align}\label{Leff}
    L=L_{\rm m}+L_{\rm em} - V_{\rm ext} - V_{\rm int}.
\end{align}
Here, the free-field Lagrangian of the scattering field $L_{\rm em}$ is defined analogous to Eq.~\eqref{freefield}, but replacing $\mathbf{A}_{\rm tot}(\mathbf{r})$ by $\mathbf{A}(\mathbf{r})$. The potential $V_{\rm ext}$ combines the optical potential and the electrostatic dipole-dipole interaction of the particles in the external electric field $\mathbf{E}_{\rm ext}(\mathbf{r},t)=-\partial_t \mathbf{A}_{\rm ext}(\mathbf{r},t)$,
\begin{align}\label{externalpotential}
    V_{\rm ext} = -\frac{\varepsilon_0}{2}&\sum_{j=1}^N V_j \mathbf{E}_{\rm ext}(\mathbf{r}_j,t)\cdot\upchi_j\mathbf{E}_{\rm ext}(\mathbf{r}_j,t)\nonumber\\ - \frac{\varepsilon_0}{2}&\sum_{\substack{j,j'=1\\ j\neq j'}}^N V_j V_{j'} \mathbf{E}_{\rm ext}(\mathbf{r}_j,t) \nonumber\\  &\cdot\upchi_j{\sf G}_0(\mathbf{r}_j - \mathbf{r}_{j'})\upchi_{j'}\mathbf{E}_{\rm ext}(\mathbf{r}_{j'},t),
\end{align}
with $V_j$ the particle volumes. The last term is the effective light-matter interaction potential
\begin{align}\label{Vint}
V_{\rm int} = - \varepsilon_0 \sum_{j=1}^N \int_{\mathcal{V}_j} d^3\mathbf{r}\, \partial_t \mathbf{E}_{\rm ext}(\mathbf{r},t)\cdot\upchi_j\mathbf{A}(\mathbf{r}),
\end{align}
describing that the polarization current $\varepsilon_0\upchi_j\partial_t\mathbf{E}_{\rm ext}(\mathbf{r},t)$, which is induced by the external laser, couples to the electromagnetic scattering field $\mathbf{A}(\mathbf{r})$.

The total time derivative of the function $S$ can be removed by means of a mechanical gauge transformation, which can be seen as performing the classical analogue of the inverse Power-Woolley-Zienau transformation on the atomic level \cite{craig1998}. Appendix \ref{appgauge} gives the details of the applied approximations to arrive at Eq.~\eqref{Leff} and the specific form of the function $S$.

Note that our assumption of ellipsoidal particles can be generalized to particles of arbitrary shape. Then, the interaction-free integral kernel ${\sf K}_0(\mathbf{r},\mathbf{r}')$ cannot be given explicitly in general. However, identifying the polarisability tensor of an ellipsoidal particle as $\upalpha_j = \varepsilon_0 V_j \upchi_j$, one can analogously define a polarisability for non-ellipsoidal particles as
\begin{align}\label{polarizability}
    \upalpha_j = \varepsilon_0\int_{\mathcal{V}_j}d^3\mathbf{r}\int_{\mathcal{V}_j}d^3\mathbf{r}' {\sf K}_0(\mathbf{r},\mathbf{r}'),
\end{align}
and still use Eq.~\eqref{Leff}. This is consistent with the Rayleigh-Gans approximation for light scattering off small dielectrics \cite{hulst1981,bohren2008}. It can be shown from Eq.~\eqref{Kimpeq} that the polarizability tensors \eqref{polarizability} do not depend on the center of mass position of the particles.

\subsection{Light-matter Hamiltonian} To derive the total Hamiltonian of the system we introduce the canonical momenta of the generalized mechanical coordinates $q_j$ as $p_j = \partial L/\partial\dot q_j$, and the conjugate momentum field as the functional derivative
\begin{align}
    \bm{\Pi}(\mathbf{r}) = \frac{\delta L}{\delta [\partial_t \mathbf{A}(\mathbf{r})]} = \varepsilon_0 \partial_t \mathbf{A}(\mathbf{r}).
\end{align}
Then, the total Hamiltonian is obtained by the Legendre transformation of $L$ as 
\begin{align}\label{hamiltonian}
    H = H_{\rm m} + H_{\rm em} + V_{\rm ext} + V_{\rm int}.
\end{align}
It involves the free particle Hamiltonian $H_{\rm m}$ as the Legendre transform of $L_{\rm m}$ and the free field Hamiltonian
\begin{align}\label{freefieldhamiltonian}
    H_{\rm em} = \int d^3\mathbf{r}\left(\frac{1}{2\varepsilon_0}\left[\bm{\Pi}(\mathbf{r}) \right]^2 + \frac{1}{2\mu_0} \left[ \nabla\times\mathbf{A}(\mathbf{r}) \right]^2\right),
\end{align}
yielding the total energy of the scattering field. The Hamiltonian \eqref{hamiltonian} can now be quantized canonically, by postulating commutation relations. For the center of mass motion they can be summarized as $[q_j,p'_{j'}] = i\hbar \delta_{jj'}\delta_{qq'}$, but we note that for degrees of freedom with a curved configuration space, such as the orientation, the commutation relations may take a more complicated form \cite{dewitt1952,stickler2016,rudolph2021}. The field commutators can be summarized as
\begin{align}\label{eq:commutator}
    \mathbf{A}(\mathbf{r})\otimes\bm{\Pi}(\mathbf{r}') - [\bm{\Pi}(\mathbf{r}')\otimes \mathbf{A}(\mathbf{r})]^T = i\hbar\delta_\perp(\mathbf{r}-\mathbf{r}').
\end{align}
Here, the transverse delta function $\delta_\perp(\mathbf{r})$ appears due to the transversality of the vector potential and the momentum field. All other commutators vanish.

\section{Quantum theory of optical binding}\label{traceout}
The quantum master equation of optical binding can now be obtained from the light-matter Hamiltonian \eqref{hamiltonian} by tracing out the electromagnetic degrees of freedom described by $\mathbf{A}(\mathbf{r})$. The external electromagnetic drive is chosen to be monochromatic, $\mathbf{E}_{\rm ext}(\mathbf{r},t)=\text{Re}[\mathbf{E}_{\rm L}(\mathbf{r})\exp(-i\omega_{\rm L} t)]$, with $\omega_{\rm L}$ the laser frequency (typically infrared) and $\mathbf{E}_{\rm L}(\mathbf{r})$ the complex laser field. The dynamical field operator  $\mathbf{A}(\mathbf{r})$ is decomposed into plane waves with wave vectors $\mathbf{k}$, transverse polarization vectors $\mathbf{t}_{\mathbf{k}s}$ ($s=1,2$ and $\mathbf{k}\cdot\mathbf{t}_{\mathbf{k}s}=0$) and electromagnetic annihilation operators $b_{\mathbf{k}s}$ defined as
\begin{align}
    b_{\mathbf{k}s} = \sqrt{\frac{\varepsilon_0 \omega_k}{2\hbar L^3}} \int d^3\mathbf{r}\,e^{-i\mathbf{k}\cdot\mathbf{r}}\mathbf{t}_{\mathbf{k}s}^* \cdot\left( \mathbf{A}(\mathbf{r}) + i\frac{\bm{\Pi}(\mathbf{r})}{\varepsilon_0\omega_k} \right),
\end{align}
with $\omega_k = ck$ and $L^3$ the quantization volume. It follows from Eq.~\eqref{eq:commutator} that $[b_{\mathbf{k}s},b_{\mathbf{k}'s'}^\dagger]=\delta_{\mathbf{k}\mathbf{k'}}\delta_{ss'}$ and $[b_{\mathbf{k}s},b_{\mathbf{k}'s'}]=0$. The vector potential and conjugate momentum field are then
\begin{subequations}
\begin{align}
    \mathbf{A}(\mathbf{r}) = \sum_{\mathbf{k}s}\left( \sqrt{\frac{\hbar}{2\varepsilon_0\omega_k L^3}} \mathbf{t}_{\mathbf{k}s} e^{i\mathbf{k}\cdot\mathbf{r}} b_{\mathbf{k}s} + \text{H.c.}\right)
\end{align}
\begin{align}
    \mathbf{\Pi}(\mathbf{r}) = \sum_{\mathbf{k}s}\left(\frac{1}{i} \sqrt{\frac{\hbar\omega_k \varepsilon_0}{2 L^3}} \mathbf{t}_{\mathbf{k}s} e^{i\mathbf{k}\cdot\mathbf{r}} b_{\mathbf{k}s} + \text{H.c.}\right),
\end{align}
\end{subequations}
so that the free field Hamiltonian \eqref{freefieldhamiltonian} reads
\begin{align}
    H_{\rm em} = \sum_{\mathbf{k}s} \hbar\omega_k \left(b_{\mathbf{k}s}^\dagger b_{\mathbf{k}s} + \frac 1 2 \right).
\end{align}

The partial trace over the electromagnetic Hilbert space is carried out in the interaction picture with respect to the free mechanical evolution and the free field energy, $H_{\rm m} + V_{\rm ext} +H_{\rm em}$. For ease of notation, we do not use a different symbol for the quantum state in the Schrödinger or interaction picture, but will denote the interaction picture versions of all other operators $A$ by $A(t)$. The Hamiltonian in the interaction picture $V_{\rm int}(t)$ follows by replacing $b_{\mathbf{k}s}$ in $V_{\rm int}$ by $b_{\mathbf{k}s}(t) = b_{\mathbf{k}s} \exp(-i\omega_k t)$ and $q_j$ by $q_j(t)$, where the latter is the time evolution of $q_j$ in absence of transverse fields.

\subsection{Born-Markov-approximation}
In order to trace out the electromagnetic field, we next perform the Born-Markov approximation for the field in the vacuum state $|0\rangle$. For this, the Schrödinger equation is integrated and iterated to the second order in the interaction $V_{\rm int}(t)$ to arrive at a coarse-grained Schrödinger equation for the total quantum state $|\psi_{\rm tot}(t)\rangle$. The change of the state $\Delta|\psi_{\rm tot}(t)\rangle = |\psi_{\rm tot}(t+\Delta t)\rangle - |\psi_{\rm tot}(t)\rangle$ during the time step $\Delta t$ then reads
\begin{align}\label{discretisedschroedinger}
    &\Delta|\psi_{\rm tot}(t)\rangle = -\frac{i}{\hbar}\int_t^{t+\Delta t} dt' V_{\rm int}(t')|\psi_{\rm tot}(t)\rangle \nonumber \\ &- \frac{1}{\hbar^2} \int_{t}^{t+\Delta t}dt' \int_{t}^{t'}dt'' V_{\rm int}(t') V_{\rm int}(t'')|\psi_{\rm tot}(t'')\rangle.
\end{align}
Choosing $\Delta t$ much smaller than the mechanical timescale and much greater than an optical period, $\omega_{\rm L}\Delta t \gg 1$, the mechanical coordinates can be approximated as $q_j(t')\approx q_j(t'') \approx q_j(t)$. In addition, we set $|\psi_{\rm tot}(t)\rangle \approx |\psi(t)\rangle\otimes|0\rangle$, where  $|\psi(t)\rangle$ denotes a pure state of the mechanical degrees of freedom. Since the electromagnetic field remains approximately in its vacuum state we can rewrite Eq.~\eqref{discretisedschroedinger} by using that $b_{\mathbf{k}s}|0\rangle = 0$ and by neglecting double photonic excitations, such as $b_{\mathbf{k}s}^\dagger b_{\mathbf{k}'s'}^\dagger |\psi_{\rm tot} (t)\rangle$, as
\begin{align}\label{discretisedschroedinger2}
    &\Delta|\psi_{\rm tot}(t)\rangle \approx \Delta B^\dagger(t) |\psi_{\rm tot}(t)\rangle + \Lambda(t)|\psi_{\rm tot}(t)\rangle.
\end{align}
The operator $\Delta B(t)$ acts both on the mechanical and the electromagnetic degrees of freedom,
\begin{align}
    \Delta B(t) =& -\sum_{j=1}^N  \int_{\mathcal{V}_j(t)} d^3\mathbf{r} \sum_{\mathbf{k}s} \sqrt{\frac{\varepsilon_0 \omega_{\rm L}^2}{8\hbar\omega_k L^3}}\int_t^{t+\Delta t}dt' \nonumber\\ & [\mathbf{E}_{\rm L}(\mathbf{r})e^{-i\omega_{\rm L} t'} - \text{c.c.}]\cdot\upchi_j(t)\mathbf{t}_{\mathbf{k}s}e^{i\mathbf{k}\cdot\mathbf{r}}b_{\mathbf{k}s}(t'),
\end{align}
with operator-valued $\mathcal{V}_j(t) \equiv \mathcal{V}_j[q_j(t)]$. The operator $\Lambda(t)$ is given by
\begin{widetext}
\begin{align}\label{lambda}
    \Lambda(t) = & \sum_{j,j'=1}^N\int_{\mathcal{V}_j(t)} d^3\mathbf{r}\int_{\mathcal{V}_{j'}(t)}d^3\mathbf{r}'\sum_{\mathbf{k}s}\frac{\varepsilon_0\omega_{\rm L}^2}{8\hbar\omega_k L^3}e^{i\mathbf{k}\cdot(\mathbf{r}-\mathbf{r}')}\int_t^{t+\Delta t}dt'\int_t^{t'}dt'' e^{-i\omega_k (t'-t'')}\nonumber\\ &[\mathbf{E}_{\rm L}(\mathbf{r})e^{-i\omega_{\rm L} t'} - \text{c.c.}]\cdot\upchi_j(t)(\mathbf{t}_{\mathbf{k}s}\otimes\mathbf{t}_{\mathbf{k}s}^*)\upchi_{j'}(t) [\mathbf{E}_{\rm L}(\mathbf{r}')e^{-i\omega_{\rm L} t''} - \text{c.c.}].
\end{align}
\end{widetext}
Since $\Lambda(t)$ acts in the mechanical subspace only, the corresponding Schr\"{o}dinger picture operator is obtained by dropping the time dependence of the mechanical degrees of freedom. For $\omega_{\rm L}\Delta t \gg 1$ two of the four integrals over $t'$ and $t''$ in \eqref{lambda} vanish while the other two can be calculated  by using \cite{gardiner2015}
\begin{align}\label{integralnotvanishing}
    &\int_t^{t+\Delta t}dt'\int_t^{t'}dt'' e^{-i(\omega_k\mp\omega_{\rm L})(t'-t'')} \nonumber\\ &\approx\pi\Delta t \delta(\omega_k \mp \omega_{\rm L}) - i\Delta t\mathcal{P}\frac{1}{\omega_k \mp \omega_{\rm L}},
\end{align}
with $\mathcal{P}$ the Cauchy principal value. The continuum limit amounts to  approximating
\begin{align}
    \frac{(2\pi)^3}{L^3}\sum_{\mathbf{k}} \approx \int d^3\mathbf{k} = \int dk k^2 \int d^2\mathbf{n},
\end{align}
where $\mathbf{k}=k\mathbf{n}$. Using that the particle sizes are much smaller than the optical wavelength, one can write Eq.~\eqref{lambda} as
\begin{align}
    \Lambda(t) \approx  -\frac{i}{\hbar}H_{\rm Lamb}(t)\Delta t -\frac{1}{2} \int d^2\mathbf{n} \sum_s L_{\mathbf{n}s}^\dagger(t)  L_{\mathbf{n}s}(t) \Delta t,
\end{align}
with the Lamb shift in the interaction picture
\begin{widetext}
\begin{align}\label{lambshift}
    H_{\rm Lamb}(t) = -\sum_{j,j'=1}^N \int_{\mathcal{V}_j(t)}d^3\mathbf{r}\int_{\mathcal{V}_{j'}(t)}d^3\mathbf{r}' \int d^3\mathbf{k}\sum_s \frac{\varepsilon_0 k_{\rm L}^2}{4(2\pi)^3} e^{i\mathbf{k}\cdot(\mathbf{r}-\mathbf{r}')} \mathbf{E}_{\rm L}^*(\mathbf{r}')\cdot\upchi_{j'}(t)(\mathbf{t}_{\mathbf{k}s}\otimes\mathbf{t}_{\mathbf{k}s}^*)\upchi_j(t)\mathbf{E}_{\rm L}(\mathbf{r})\mathcal{P}\frac{1}{k^2-k_{\rm L}^2}.
\end{align}
\end{widetext}
The operators
\begin{align}\label{lindbladians}
    L_{\mathbf{n}s}(t) = \sum_{j=1}^N \sqrt{\frac{\varepsilon_0 k_{\rm L}^3}{2\hbar}}\frac{V_j}{4\pi} \mathbf{t}_{\mathbf{n}s}^*\cdot\upchi_j(t)\mathbf{E}_{\rm L}[\mathbf{r}_j(t)] e^{-ik_{\rm L}\mathbf{n}\cdot\mathbf{r}_j(t)}
\end{align}
enact the momentum kick associated with a single photon scattering event. Here we denote the laser wave number by $k_{\rm L} = \omega_{\rm L}/c$ and the scattered photon polarizations by $\mathbf{t}_{\mathbf{n}s}$.

\subsection{Conservative part of optical binding}

Next we demonstrate that the Lamb shift \eqref{lambshift} yields the conservative optical binding interaction due to light scattering, which adds to the electrostatic coupling in \eqref{externalpotential}. This radiative contribution to optical binding, which is crucial to recover the classical interparticle coupling \cite{rieser2022}, requires treating the light-matter interaction according to Eq.~\eqref{Vint}. 

We simplify Eq.~\eqref{lambshift} by using the transverse completeness of the polarization vectors, $\sum_s \mathbf{t}_{\mathbf{k}s}\otimes\mathbf{t}_{\mathbf{k}s}^* = \mathbb{1}-\mathbf{k}\otimes\mathbf{k}/k^2$, so that
\begin{align}\label{completeness}
    \int d^3 \mathbf{k }\sum_{s} \mathbf{t}_{\mathbf{k}s}\otimes \mathbf{t}_{\mathbf{k}s}^* e^{i {\bf k}\cdot {\bf r}} f({\bf k}) = &\, [\mathbb{1} - (\nabla\otimes\nabla) \Delta^{-1}]\nonumber \\
& \times \int d^3\mathbf{k} e^{i {\bf k}\cdot {\bf r}} f({\bf k}),
\end{align}
for arbitrary $f({\bf k})$. Here $\Delta^{-1}$ is defined through its Fourier transform $[\Delta^{-1}f](\mathbf{k}) = -f(\mathbf{k})/k^2$, so that in position space one has
\begin{align}
    [\Delta^{-1}f](\mathbf{r}) = -\int d^3\mathbf{r}' \frac{f(\mathbf{r}')}{4\pi|\mathbf{r}-\mathbf{r}'|}.
\end{align}
For natural boundary conditions $\Delta^{-1}$ is thus the inverse Laplacian. In addition, we note that
\begin{equation}
    \mathcal{P}\frac{1}{k^2-k_{\rm L}^2} = \text{Re}\left  (\frac{1}{k^2-k_{\rm L}^2 - i\eta}\right ),
\end{equation}
with infinitesimal $\eta>0$.

After pulling the real part to the front of the ${\bf k}$-integration in Eq.~\eqref{lambshift}, the Fourier transform can be carried out
\begin{align}\label{residue}
    \frac{k_{\rm L}^2}{(2\pi)^3}\int d^3\mathbf{k}  e^{i\mathbf{k}\cdot\mathbf{r}} \frac{1}{k^2-k_{\rm L}^2 - i\eta} = \frac{k_{\rm L}^2 e^{i k_{\rm L} r}}{4\pi r}.
\end{align}
The action of $\Delta^{-1}$ to the right-hand side of this expression can be evaluated by applying $\Delta^{-1}$ to $(\Delta + k_{\rm L}^2)\exp(i k_{\rm L} r)/4\pi r = -\delta(\mathbf{r})$ from the left and using the inverse of $\Delta (1/4\pi r) = -\delta(\mathbf{r})$. This yields
\begin{align}
    \Delta^{-1}\frac{k_{\rm L}^2 e^{ik_{\rm L} r}}{4\pi r} = \frac{1}{4\pi r} - \frac{e^{ik_{\rm L} r}}{4\pi r}.
\end{align}
Thus, one finally obtains
\begin{align}\label{transverseGreen}
    [\mathbb{1} - (\nabla\otimes\nabla) \Delta^{-1}] \frac{k_{\rm L}^2 e^{ik_{\rm L} r}}{4\pi r} = {\sf G}(\mathbf{r}) - {\sf G}_0(\mathbf{r}),
\end{align}
with the full electromagnetic and electrostatic dipole Green tensors \eqref{greentensor} and \eqref{staticgreen}, respectively. The Lamb shift in the interaction picture thus takes the form
\begin{widetext}
\begin{align}\label{lambshift2}
    H_{\rm Lamb}(t) = -\frac{\varepsilon_0 }{4}\sum_{j,j'=1}^N \int_{\mathcal{V}_j(t)}d^3\mathbf{r}\int_{\mathcal{V}_{j'}(t)}d^3\mathbf{r}'  \mathbf{E}_{\rm L}^*(\mathbf{r}')\cdot\upchi_{j'}(t){\rm Re}\left [{\sf G}({\bf r} - {\bf r}') - {\sf G}_0({\bf r} - {\bf r}')\right ]\upchi_j(t)\mathbf{E}_{\rm L}(\mathbf{r}).
\end{align}
The integrals over the particle volumes can be carried out for  particles small in comparison with the laser wavelength and their separation. For $j=j'$, the transverse Green function (\ref{transverseGreen}) has to be approximated up to the third order in $k_{\rm L}$ as
\begin{align}\label{approxGreen}
    {\sf G}(\mathbf{r})-{\sf G}_0(\mathbf{r}) \approx \frac{k_{\rm L}^2}{8\pi} \frac{r^2\mathbb{1} + \mathbf{r}\otimes\mathbf{r}}{r^3} + i\frac{k_{\rm L}^3}{6\pi}\mathbb{1}.
\end{align}
Transforming back to the Schrödinger picture, this yields the Lamb shift
\begin{align}\label{lambfinal}
H_{\rm Lamb} =& -\frac{\varepsilon_0}{4}\sum_{j=1}^N V_j \mathbf{E}_{\rm L}^*(\mathbf{r}_j)\cdot\delta\upchi_j\mathbf{E}_{\rm L}(\mathbf{r}_j)  - \frac{\varepsilon_0}{4}\sum_{\substack{j,j'=1\\ j\neq j'}}^N V_j V_{j'} \mathbf{E}_{\rm L}^*(\mathbf{r}_{j'})\cdot\upchi_{j'}\text{Re}[{\sf G}(\mathbf{r}_j-\mathbf{r}_{j'})-{\sf G}_0(\mathbf{r}_j-\mathbf{r}_{j'})]\upchi_j \mathbf{E}_{\rm L}(\mathbf{r}_j),
\end{align}
with the radiation correction to the susceptibility tensor \cite{rudolph2021}
\begin{align}\label{deltachi}
    \delta\upchi_j = \frac{k_{\rm L}^2}{8\pi V_j} \int_{\mathcal{V}_j}d^3\mathbf{r}\int_{\mathcal{V}_j}d^3\mathbf{r}' \upchi_j \frac{|\mathbf{r}-\mathbf{r}'|^2\mathbb{1} + (\mathbf{r}-\mathbf{r}')\otimes(\mathbf{r}-\mathbf{r}')}{|\mathbf{r}-\mathbf{r}'|^3}\upchi_j,
\end{align}
\end{widetext}
which can be shown to be independent of the particle positions. Adding the Lamb shift \eqref{lambfinal} to the electrostatic optical binding interaction \eqref{externalpotential} shows that the free Green tensor cancels out such that the conservative interaction is determined by the full electromagnetic Green tensor \eqref{greentensor}. The conservative part of the interaction thus exhibits retardation effects due to the finite speed of light. The same interaction  is  obtained in the classical treatment \cite{rieser2022}, based on integrating out Maxwell's stress tensor. We will see next that the Lamb shift \eqref{lambfinal} causes the conservative part of optical binding (apart from the near-field contribution ${\sf G}_0$), while the operators \eqref{lindbladians} describe the nonconservative part.

\subsection{Optical binding master equation}

We can now derive the quantum master equation of optical binding by reformulating Eq.~\eqref{discretisedschroedinger2} in terms of the density operator and tracing out the electromagnetic field. The temporal increment of the reduced state
\begin{align}
    \Delta\rho(t) = \, &\text{tr}_{\rm em}\lbrace |\psi_{\rm tot}(t+\Delta t)\rangle\langle \psi_{\rm tot}(t+\Delta t)| \rbrace \nonumber\\ &- |\psi(t)\rangle \langle \psi(t)|,
\end{align}
then follows from Eq.~\eqref{discretisedschroedinger2} as
\begin{align}\label{partialtrace}
    \Delta\rho(t) =\, & [\Lambda(t)\rho(t) + \rho(t)\Lambda^\dagger(t)]\Delta t \nonumber\\ &+  \text{tr}_{\rm em} \lbrace \Delta B^\dagger(t) |\psi_{\rm tot}(t)\rangle\langle \psi_{\rm tot}(t)|\Delta B(t) \rbrace,
\end{align}
where we used that the terms linear in $\Delta B(t)$ and $\Delta B^\dagger(t)$ vanish. Since $\text{tr}_{\rm em} \lbrace b_{\mathbf{k}'s'}^\dagger|\psi_{\rm tot}(t)\rangle\langle\psi_{\rm tot}(t)|b_{\mathbf{k}s} \rbrace = \delta_{\mathbf{k}\mathbf{k}'}\delta_{ss'}\rho (t)$, the term quadratic in $\Delta B(t)$ evaluates to
\begin{align}
    &\text{tr}_{\rm em} \lbrace \Delta B^\dagger(t) |\psi_{\rm tot}(t)\rangle\langle \psi_{\rm tot}(t)|\Delta B(t) \rbrace = \nonumber\\ & \int d^2\mathbf{n}\sum_s  L_{\mathbf{n}s}(t) \rho (t) L_{\mathbf{n}s}^\dagger(t) \Delta t,
\end{align}
where we made the same approximations as in Eq.~\eqref{lambda}.
Switching back to the Schrödinger picture, averaging the external potential\eqref{externalpotential} over one optical cycle, and taking the limit $\Delta t\rightarrow 0$, the optical binding master equation is finally obtained as
\begin{align}\label{master}
    \partial_t \rho =& -\frac{i}{\hbar}[H_{\rm m} + V_{\rm L} + V_{\rm opt},\rho]\nonumber\\ &+ \int d^2\mathbf{n}\sum_s \left( L_{\mathbf{n}s}\rho L_{\mathbf{n}s}^\dagger - \frac 1 2 \lbrace L_{\mathbf{n}s}^\dagger L_{\mathbf{n}s},\rho \rbrace \right).
\end{align}
It involves the time-averaged optical potential
\begin{align}\label{opticalpotential}
    V_{\rm L} = -\frac{\varepsilon_0}{4}\sum_{j=1}^N V_j \mathbf{E}_{\rm L}^*(\mathbf{r}_j)\cdot\tilde{\upchi}_j\mathbf{E}_{\rm L}(\mathbf{r}_j)
\end{align}
featuring the renormalized susceptibility tensors \begin{equation}
    \tilde{\upchi}_j=\upchi_j + \delta\upchi_j
\end{equation}
of the particles. Note that the radiation correction $\delta\upchi_j$, Eq.~\eqref{deltachi}, scales with $V_j^{2/3}$, consistent with the classical calculation \cite{rudolph2021}.

The conservative part of the optical binding interaction
\begin{align}\label{conservativebinding}
    V_{\rm opt} =  -\frac{\varepsilon_0}{4} \sum_{\substack{j,j'=1\\ j\neq j'}}^N & V_j V_{j'} \mathbf{E}_{\rm L}^*(\mathbf{r}_{j'})\nonumber\\ &\cdot\upchi_{j'}\text{Re}[{\sf G}(\mathbf{r}_j-\mathbf{r}_{j'})]\upchi_j\mathbf{E}_{\rm L}(\mathbf{r}_j),
\end{align}
depends on the real part of the electromagnetic dipole Green tensor. This expression can be interpreted as the potential energy of two interacting induced dipoles. This conservative interaction is accompanied by the nonconservative optical binding interaction described by the Lindblad operators
\begin{align}\label{lindbladians2}
    L_{\mathbf{n}s} = \sum_{j=1}^N \sqrt{\frac{\varepsilon_0 k_{\rm L}^3}{2\hbar}}\frac{V_j}{4\pi} \mathbf{t}_{\mathbf{n}s}^*\cdot\upchi_j\mathbf{E}_{\rm L}(\mathbf{r}_j) e^{-ik_{\rm L}\mathbf{n}\cdot\mathbf{r}_j}.
\end{align}
They can be viewed as the coherent sum of the single-particle scattering amplitudes of all particles \cite{rudolph2021}. It gives rise to interference between the photon scattering amplitudes off different particles, as is also the case in superradiance \cite{lehmberg1970,lehmberg1970part2,agarwal1970,agarwal1971,gross1982,vogt1996}. The interference leads to nonreciprocal coupling (see below), in addition to the nonconservative radiation pressure forces and decoherence present also for single particles \cite{rudolph2021}. We note that the optical binding master equation \eqref{master} also applies to far-detuned atomic systems, as treated in \cite{vogt1996,maximo2018optical,shahmoon2020}, if one replaces the induced dipole moment by the dipole operator. The master equation \eqref{master} is then recovered when the electronic degrees of freedom are eliminated adiabatically, so that the the optical response of the atom is characterized by its polarizability \cite{vogt1996}.

The optical binding master equation in \eqref{master} is fully consistent with the nonreciprocal classical equations of motion obtained in \cite{dholakia2010,rieser2022} and in Sec.~\ref{classicalbinding}, as can be checked by first deriving the equations of motion for the position and momentum expectation values from \eqref{master} and then replacing all operators by their expectation values. We note that the classical equations of motion can also be obtained from the quantum Langevin equations, which are equivalent to \eqref{master}-\eqref{lindbladians2}. The latter are derived in Appendix \ref{applangevin}.

\section{Non-Hermitian quantum arrays}\label{deeplytrapped}
One of the central features of optical binding forces is their inherent nonreciprocity. To shed light on the effect of optical binding in multiparticle levitated optomechanics, we investigate the situation where multiple nanospheres are deeply trapped in optical tweezers, such that the interaction can be expanded harmonically. This yields the theoretical toolbox required to understand future experiments with co-levitated nanoparticles similar to \cite{rieser2022}, but in the deep quantum regime. Furthermore, we will study the implications of quantum optical binding for upcoming quantum experiments with optically interacting nanoparticle arrays.

\subsection{multiparticle array} \label{sec:array} We focus in the following on rigid spheres characterized by a homogeneous dielectric constant $\varepsilon$. For such particles, the susceptibility tensor Eq.~\eqref{eq:suscepttsensor}  is isotropic ($\upchi_j = \chi\mathbb{1}$ and $\tilde{\upchi}_j = \tilde{\chi}\mathbb{1}$) so that rotations can be traced out from the dynamics as they only enter via the orientational dependence of the susceptibility tensor $\upchi_j$. Further, we consider all spheres in the array to have the same susceptibility, but we allow for different volumes $V_j$ and masses $m_j$.

We take the tweezer for each particle to have the same progagation direction $\mathbf{e}_z$, and their foci $\mathbf{d}_j$ to be located on an orthogonal plane, $\mathbf{e}_z \cdot\mathbf{d}_j = 0$. Moreover, we assume the tweezers to have identical waists $w$ and Rayleigh ranges $z_{\rm R}=k_{\rm L} w^2/2$, but different field strength maxima $\mathbf{E}_j$ to control the local trapping frequencies. The laser field can thus be written as \cite{rudolph2021}
\begin{align}
    \mathbf{E}_{\rm L}(\mathbf{r}) = \sum_{j=1}^N \mathbf{E}_j e^{ik_{\rm L} z} f_{\rm tw}(\mathbf{r}-\mathbf{d}_j),
\end{align}
with the tweezer field envelope
\begin{align}
    f_{\rm tw}(\mathbf{r}) = \frac{1}{1+iz/z_{\rm R}} \exp\left( -\frac{x^2 + y^2}{w^2(1+iz/z_{\rm R})} \right).
\end{align}

The beam waist is typically much smaller than the corresponding Rayleigh range, so that the radial trapping frequencies are far detuned from that for the motion along the optical axis. Since we assume the coordinates of the particles transverse to the beam propagation direction to be deeply trapped, we can safely ignore the transverse degrees of freedom and focus on the $z$-motion by replacing $\mathbf{r}_j = \mathbf{d}_j + z_j \mathbf{e}_z$. 

\subsubsection{Master equation}

The total kinetic energy of the particles is $H_{\rm m} = \sum_j p_j^2/2m_j$, with $p_j$ the momentum operators for motion along the optical axis. Additionally, we assume that all particles stay near the foci of their respective tweezers, $|z_j| \ll z_{\rm R}$, and that the distance between all tweezers is much greater than the beam waist, so that the tweezers do not overlap (they still influence distant particles via optical binding). Then, for small deviations from the tweezer foci at $z_j=0$, the potential energy due to the laser beams is approximately
\begin{align}
    V_{\rm L} \approx -\frac{\varepsilon_0\tilde{\chi}}{4}\sum_{j=1}^N V_j |\mathbf{E}_j|^2\left( 1 - \frac{z_j^2}{z_{\rm R}^2} \right),
\end{align}
from which we identify the particle trapping frequencies via $m_j \omega_j^2 = \varepsilon_0 \tilde{\chi} V_j |\mathbf{E}_j|^2/2 z_{\rm R}^2$.

Next, the optical binding potential \eqref{conservativebinding} is harmonically expanded around $z_j=0$ by approximating the laser beam near the respective tweezer focus by $\mathbf{E}_{\rm L}(\mathbf{r}_j)\approx \mathbf{E}_j\exp[i(k_{\rm L}-1/z_{\rm R})z_j]$, where the local effective wave number is reduced by $1/z_{\rm R}$ due to the Gouy phase \cite{gonzalez2019,rudolph2021}. Defining the distance between two tweezer foci as $d_{jj'}=|\mathbf{d}_j-\mathbf{d}_{j'}|$ and the respective connecting vector as $\mathbf{n}_{jj'}=(\mathbf{d}_j-\mathbf{d}_{j'})/d_{jj'}$, the harmonically approximated optical binding potential reads
\begin{align}
 V_{\rm opt} \approx & -\sum_{\substack{j,j'=1\\ j\neq j'}}^N  \frac{\varepsilon_0 \chi^2 k_{\rm L}^2 V_j V_{j'}}{16\pi d_{jj'}} \mathbf{E}_{j'}^*\cdot\left(\mathbb{1} - \mathbf{n}_{jj'}\otimes\mathbf{n}_{jj'}\right)\mathbf{E}_j\nonumber\\ & \times \cos(k_{\rm L} d_{jj'}) \Biggl[ 1 + i\left(k_{\rm L}-\frac{1}{z_{\rm R}}\right)(z_j-z_{j'}) \nonumber\\ &- \left(k_{\rm L}-\frac{1}{z_{\rm R}}\right)^2\frac{(z_j-z_{j'})^2}{2} \Biggr].
\end{align}
Note that here we assume  the particles to interact predominantly via their scattered fields in the far field, implying that all contributions of order higher than $1/d_{jj'}$ are negligible. Therefore, only the far-field contribution to the Green tensor \eqref{greentensor} evaluated at the tweezer foci contributes.

The Lindblad operators \eqref{lindbladians2} are expanded to quadratic order in the position operators $z_j$,
\begin{align}
    L_{\mathbf{n}s} \approx & \sum_{j=1}^N \sqrt{\frac{\varepsilon_0 k_{\rm L}^3}{2\hbar}} \frac{V_j\chi}{4\pi}\mathbf{t}_{\mathbf{n}s}^*\cdot\mathbf{E}_j \Biggl[ 1 +i\left( k_{\rm L} - \frac{1}{z_{\rm R}} - k_{\rm L} n_z \right)z_j \nonumber\\ &-\left( k_{\rm L} - \frac{1}{z_{\rm R}} - k_{\rm L} n_z \right)^2 \frac{z_j^2}{2} \Biggr],
\end{align}
with $n_z = \mathbf{n}\cdot\mathbf{e}_z$ the $z$-component of the photon scattering direction.

\begin{figure*}[t]
	\centering
	\includegraphics[width=1\linewidth]{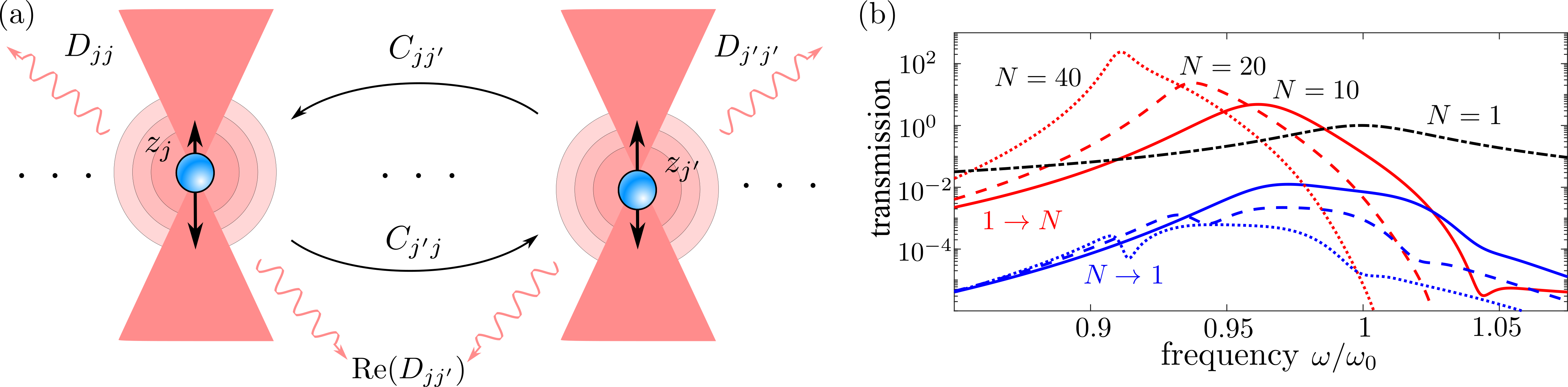}
    \caption{(a) An array of harmonically trapped nanoparticles with positions $z_j$ along the optical axis, driven by the same laser. The light scattered off the particles couples their motion nonreciprocally with coupling constants $C_{jj'}$, and imprints photon shot noise, which leads to recoil heating with diffusion constants $D_{jj}$. The photon shot noise is correlated, which is described by Re($D_{jj'}$). (b) Absolute square of the mechanical susceptibilities $\chi_{N1}[\omega]$ (red) and $\chi_{1N}[\omega]$ (blue), see Eq.~\eqref{eq:susceptibility}, in a linear chain of $N=10$ (solid line), $N=20$ (dashed) and $N=40$ (dotted) particles. The dotted-dashed, black line is the susceptibilty of a single particle. The next-neighbor distance in the chain is always $d_{\rm next}=(2\pi n + \pi/4)/k_{\rm L}$, and the next-neighbor tweezer phase difference $\varphi_{\rm next} = \pi/4$. The trapping frequency $\omega_0 = 20\gamma_{\rm g}$ and the coupling rate $g=\gamma_{\rm g}$ are state of the art for the case of two particles, see \cite{rieser2022}.}
	 \label{fig:sketchmulti}
\end{figure*}

Inserting these expressions into the master equation (\ref{master}) and evaluating the integrals to first order in $1/d_{jj'}$ yields the optical binding master equation for small displacements along the beam propagation direction,
\begin{align}\label{masterlinear}
    \partial_t \rho = &-\frac{i}{\hbar} [H_{\rm na},\rho] \nonumber\\ &+ \sum_{j,j' = 1}^N \frac{2D_{jj'}}{\hbar^2} \left( z_j \rho z_{j'} - \frac{1}{2} \lbrace z_{j'} z_j,\rho \rbrace \right).
\end{align}
Here, the Hamiltonian of the nanoparticle array takes the form
\begin{align}\label{eq:hamiltonianna}
    H_{\rm na} = & \sum_{j=1}^N  \left(  \frac{p_j^2}{2m_j} +\frac{1}{2} \left (m_j\omega_j^2 + K_j\right )z_j^2 - F_j z_j\right)\nonumber\\ & - \sum_{\substack{j,j'=1\\ j\neq j'}}^N \frac{C_{jj'}}{2} z_j z_{j'}.
\end{align}
Apart from a renormalization of the trapping frequencies determined by
\begin{align}
    K_j =& \sum_{\substack{j'=1\\ j'\neq j}}^N C_{jj'},
\end{align}
it describes a linear interaction between the nanoparticles with  coupling constants
\begin{align}\label{couplings}
    C_{jj'} =& \frac{\varepsilon_0 \chi^2 V_j V_{j'}k_{\rm L}^2 (k_{\rm L}-1/z_{\rm R})^2}{8\pi d_{jj'}}\nonumber\\ &\times\text{Re}\left[ e^{ik_{\rm L} d_{jj'}} \mathbf{E}_j^* \cdot(\mathbb{1}-\mathbf{n}_{jj'}\otimes\mathbf{n}_{jj'})\mathbf{E}_{j'} \right].
\end{align}
Note that only the symmetric part $C_{jj'} + C_{j'j}$ contributes to  \eqref{eq:hamiltonianna}. Moreover, each particle experiences a constant force 
\begin{align}
    & F_j = \frac{\varepsilon_0 \chi^2 V_j k_{\rm L}^2(k_{\rm L}-1/z_{\rm R})}{8\pi}\Biggl[\frac 2 3 V_j k_{\rm L}|\mathbf{E}_j|^2  \nonumber\\ & +\sum_{\substack{j'=1\\ j'\neq j}}^N \frac{V_{j'}}{d_{jj'}}\text{Im}\left[ e^{ik_{\rm L} d_{jj'}}\mathbf{E}_j^* \cdot(\mathbb{1}-\mathbf{n}_{jj'}\otimes\mathbf{n}_{jj'})\mathbf{E}_{j'} \right]\Biggr]
\end{align}
due to the nonconservative radiation pressure exerted by the local tweezers \cite{rudolph2021} and due to a contribution from the optical interaction between the particles; this gives rise to a constant, small displacement.

The incoherent part of the time evolution \eqref{masterlinear} is described by the diffusion matrix $D_{jj'}$, with diagonal elements
\begin{subequations}\label{diffusion}
\begin{align}\label{diffusiona}
    D_{jj} = \frac{\hbar \varepsilon_0 \chi^2 V_j^2 k_{\rm L}^3 |\mathbf{E}_j|^2}{120\pi} [5(k_{\rm L}-1/z_{\rm R})^2 + 2k_{\rm L}^2],
\end{align}
and
\begin{align}\label{diffusionb}
    D_{jj'} =& \frac{\hbar\varepsilon_0 \chi^2 V_j V_{j'} k_{\rm L}^2 (k_{\rm L}-1/z_{\rm R})^2 }{16\pi d_{jj'}} \sin(k_{\rm L} d_{jj'})\nonumber\\ &\mathbf{E}_{j'}^* \cdot (\mathbb{1} - \mathbf{n}_{jj'}\otimes\mathbf{n}_{jj'})\mathbf{E}_j
\end{align}
\end{subequations}
for $j\neq j'$. This matrix is Hermitian and positive 
(as implied by $D_{jj}D_{j'j'} > |D_{jj'}|^2$), guaranteeing the complete positivity of the time evolution \eqref{masterlinear}. 
 
The diffusion matrix accounts for three distinct effects, see Fig.~\ref{fig:sketchmulti} (a): (i) The diagonal elements \eqref{diffusiona} describe  recoil heating of each individual particle due to the shot noise of the local tweezer \cite{rudolph2021,rudolph2022}, which  also occurs for non-interacting particles. (ii) The real part of the off-diagonals \eqref{diffusionb} describes correlations between the recoil noise experienced by different particles. It is a consequence of the finite overlap of the electromagnetic modes into which different particles scatter \cite{rudolph2022,gonzalez2023}. (iii) The imaginary part of $D_{jj'}$ describes a coupling between the particles $j$ and $j'$ where the principle of {\it action equals reaction} is maximally violated (antireciprocal coupling).

Experimentally, shot noise correlations can be expected to become relevant when recoil heating dominates all other noise sources. In state-of-the-art experiments with tweezer levitated particles, this occurs at pressures below $10^{-7}$\,mbar. Note that even though the impact of shot noise can be characterized by a diffusion process, the above quantum description is required for quantifying its strength.
 
That the total optical interaction may seemingly violate Newton's third law is a direct consequence of the fact that optically induced interactions are mediated via a common photonic environment, which carries away or adds momentum and energy.

\subsubsection{Quantum Langevin equations}
The quantum dynamics described by the master equations \eqref{master} and \eqref{masterlinear} can be reformulated in terms of Langevin equations. Harmonically approximating the general quantum Langevin equation (Appendix \ref{applangevin}) one obtains the linearized equations,
\begin{subequations}\label{langevinlinear}
\begin{align}
    \dot z_j =& \,\frac{p_j}{m_j}
\\
    \dot p_j = &-m_j\omega_j^2 z_j + F_j + \xi_{j} \nonumber\\ &+ \sum_{\substack{j'=1\\ j'\neq j}}^N C_{jj'}(z_{j'} - z_j),
\end{align}
\end{subequations}
which are equivalent to the linearized master equation \eqref{masterlinear}.

Importantly, the operator-valued noise forces $\xi_{j}$ associated with the different particles are correlated,
\begin{align}\label{correlator}
    \langle \xi_{j'}(t') \xi_{j}(t) \rangle = 2D_{jj'} \delta(t-t').
\end{align}
The correlators of the noise forces are in general complex with $D_{jj'} = D_{j'j}^*$, implying that $\xi_j$ and $\xi_{j'}$ do not commute,
\begin{align}
    [\xi_{j'}(t'), \xi_j(t)] = 4i\text{Im}(D_{jj'}) \delta(t-t')
\end{align}
The correlator \eqref{correlator} ensures that the (equal time) canonical commutation relations  $[z_j,p_{j'}]=i\hbar \delta_{jj'}$ are preserved under the dynamics (since $C_{jj'} - C_{j'j} = 4 {\rm Im}(D_{jj'})/\hbar$ implies that the noise correlations exactly cancel the nonreciprocal interactions). This would not be the case under nonreciprocal couplings and uncorrelated noises. The real part of the noise force correlations describes statistical correlations between the photon recoils of different particles.

The coupling constants \eqref{couplings} appearing in \eqref{langevinlinear} combine both the reciprocal and antireciprocal interactions. Importantly, $C_{jj'}$ and $C_{j'j}$ can be tuned continuously via the relative tweezer phases and distances. We note that for weak interactions, the nonreciprocity in Eq.~\eqref{langevinlinear} turns into the linearized Hatano-Nelson dimer model \cite{martello2023}.

The master equation \eqref{masterlinear}, or equivalently the quantum Langevin equations (\ref{langevinlinear}), provide the theoretical basis for describing optically interacting nanoparticles in the quantum regime. Classically, large nanoparticle arrays exhibit rich behaviour, including the non-Hermitian skin effect and non-Hermitian topological phase transitions \cite{yokomizo2023}. We note that arrays of trapped atoms can be described by quantum Langevin equations similar to Eq.~\eqref{langevinlinear} \cite{shahmoon2020}; when coupled via dissipative cavities, two-level atoms can also exhibit non-Hermitian effects \cite{joshi2023,roccati2024}. 

In the following we present some consequences and signatures of quantum optical binding in the context of unidirectional transport.
\subsection{Unidirectional transport}

For the case of two particles, $N=2$, the optical binding interaction between them can be set unidirectional, such that one particle influences the other but not vice versa.  Unidirectionality can serve as technological resource, such as for non-Hermitian quantum sensing \cite{lau2018,mcdonald2020} or for directional amplification in chains of multiple oscillators \cite{wanjura2020,wanjura2021}. To see how optical binding can give rise to unidirectional interactions we place two linearly polarized tweezers along the ${\bf e}_x$ direction at distance $d$ and denote the relative tweezer phase by $\varphi$, so that ${\bf E}_1 = {\bf E}_2 e^{i\varphi}$. Unidirectional coupling can be achieved by setting $k_{\rm L} d = \varphi = \pi/4$, so that the scattering fields  interfere destructively in one direction and constructively in the opposite one. As a consequence, the coupling constants take the form 
\begin{subequations}
\begin{align}
    C_{12} = & \, C \\
    C_{21} = & \,  0
\end{align}    
\end{subequations}
with
\begin{align}
    C = \frac{\varepsilon_0 \chi^2 V_1 V_2 k_{\rm L}^2 (k_{\rm L}-1/z_{\rm R})^2 }{8\pi d} |{\bf E}_1| |{\bf E}_2| \cos\Theta_1 \cos\Theta_2.
\end{align}
Here, $\pi/2 - \Theta_{1,2}$ is the angle between the tweezer polarization and the tweezer connecting axis. The cross diffusion coefficient is given by 
\begin{equation}
    D_{12}=(1+i)\frac{\hbar C }{4},
\end{equation}
so that the master equation \eqref{masterlinear} then becomes
\begin{align}\label{masteruni}
    \partial_t \rho =& -\frac{i}{\hbar}[H_{\rm uc},\rho] + \sum_{j,j'=1}^2 \frac{2 D_{jj'}^{\rm uni}}{\hbar^2} \left( z_j \rho z_{j'} - \frac{1}{2}\lbrace z_j z_{j'},\rho \rbrace\right)\nonumber\\ &+ \frac{C}{2\hbar} \Bigl[ i[z_1, \{z_2,\rho\}] + \mathcal{D}[z_1]\rho + \mathcal{D}[z_2]\rho \Bigr],
\end{align}
with the real-valued diffusion constants $D_{12}^{\rm uni} = D_{21}^{\rm uni} = \hbar C/4$ and $D_{jj}^{\rm uni} = D_{jj} - \hbar C/4$, and where $\mathcal{D}[A]\cdot = A\cdot A^\dagger - \{A^\dagger A ,\cdot \}/2$. Here, we defined the uncoupled particle Hamiltonian
\begin{align}\label{eq:Huc}
    H_{\rm uc} = \sum_{j=1}^2  \left(  \frac{p_j^2}{2m_j} +\frac{1}{2} \left (m_j\omega_j^2 + K_j\right )z_j^2 - F_j z_j\right).
\end{align}
The master equation \eqref{masteruni} decomposes into two parts: (i) The first line describes unitary and uncoupled dynamics of the two particles through the Hamiltonian \eqref{eq:Huc} as well as correlated quantum noise on the two particles. (ii) The second line features the standard form of unidirectional transport between quantum systems \cite{metelmann2017,clerk2022}.

One might expect that the unidirectional transport between two coupled particles can be readily extended to larger arrays of $N>2$ nanoparticles by neglecting couplings beyond the next few neighbors and using the same choice of relative tweezer phases as above. However, such systems are described by the linearized master equation \eqref{masterlinear} or the quantum Langevin equations \eqref{langevinlinear}, where the inter-particle distances $d_{jj'}$ and the complex tweezer amplitudes ${\bf E}_j$ determine the all-to-all couplings \eqref{couplings}. While the nearest-neighbor couplings in a linear chain of nanoparticles can in principle be tuned independently, one must account for the fact that the optical binding interaction is long-range, $C_{jj'}\propto 1/d_{jj'}$, see Eq.~\eqref{couplings}. The dynamics are correctly described only if the coupling of each particle to all other particles is taken into account. Moreover, the couplings between far-separated particles cannot be tuned independently of the nearest-neighbor interaction. The important consequences for particle arrays are: 

\begin{enumerate}
\item[(I)] It is impossible to build a unidirectional chain, where the transport of excitations takes place in one direction only, by positioning nanoparticles equidistantly on a line. Instead, tuning the nearest-neighbor couplings unidirectional by choosing the nearest-neighbor tweezer phase difference as $\pi/4$ and the distances as $(2\pi n + \pi/4)/k_{\rm L}$ with $n\in\mathbb{N}$, the coupling constants are always positive in one direction, while alternating between zero, positive, and negative values in the other. Directional amplification is still possible in such chains, as we show below, which enables signal enhancement in arrays of nanoparticles \cite{lau2018,porras2019,wanjura2020}. 
\item[(II)] The long-range physical arrangement of the particles matters because even distant particles influence each other. For instance, a circular arrangement of the particles is not describable by a one-dimensional chain with periodic boundary conditions.
\item[(III)] Inter-particle correlations in the thermal state of a particle array may be strongly influenced by recoil noise correlations, since they can be present even in the absence of coupling.
\item[(IV)] Paradigmatic phenomena of non-hermiticity in particle arrays, such as the non-Hermitian skin effect \cite{yao2018,bergholtz2021}, may be modified given that the edges of the particle array may still interact directly. In fact, the concepts of edges and bulk are no longer fully applicable in the presence of long-range interaction, with potentially far-reaching consequences for the non-Hermitian bulk-boundary correspondence \cite{bergholtz2021}.
\end{enumerate}

To illustrate these implications, we now show that directional amplification is indeed possible in the finite linear chains described in point (I) while their bulk limit is not well defined. In the following, we assume that all particles in this chain are identical, that their renormalized spring constants (see Eq.~\eqref{eq:hamiltonianna}) are equal, $\omega_j^2 + K_j/m = \omega_0^2$, and that all tweezers are equally polarized orthogonal to the particle chain. Adding gas damping with equal rate $\gamma_{\rm g}$ for all particles to the quantum Langevin equations \eqref{langevinlinear} yields
\begin{align}\label{eq:langevin2}
    \ddot{z}_j + \gamma_{\rm g} \dot{z}_j + \omega_0^2 z_j - \sum_{j'=1}^N\frac{C_{jj'}}{m} z_{j'}  = \frac{\xi_j}{m},
\end{align}
where we omitted the forces $F_j$ since they can be absorbed by a small shift of the particle equilibrium positions. The coupling constants can be written as
\begin{align}
    C_{jj'} = \frac{2m\omega_0 g}{|j-j'|} \cos[kd_{\rm next}|j-j'| - \varphi_{\rm next}(j-j')],
\end{align}
with $d_{\rm next} = (2\pi n + \pi/4)/k_{\rm L}$ the next-neighbor distance and $\varphi_{\rm next} = \varphi_j - \varphi_{j+1} = \pi/4$ the next-neighbor phase difference, see (I). The maximal coupling rate reads
\begin{align}\label{eq:couplingrate}
    g = \frac{\varepsilon_0 \chi^2 V^2 k_\text{L}^2(k_\text{L} - 1/z_\text{R})^2}{16\pi m\omega_0 d_\text{next}}|\mathbf{E}_0|^2,
\end{align}
where we took the maximum field strengths $|\mathbf{E}_0|$ of all tweezers in the coupling constants to be equal, which is well allowed for weak inter-particle coupling and small mechanical frequency differences. 

We note that the bulk limit is not well defined. This can be seen by noting that the ansatz of translation-invariant solutions $z_j(t) = e^{i (\kappa j - \omega_\kappa t)}$ with wave number $\kappa$ yields a dispersion relation that does not converge absolutely,
\begin{align}
    \omega_\kappa^2 = \omega_0^2 - 2g\omega_0 \sum_{j=1}^\infty \left[ \frac{1}{j}e^{-i\kappa j} + \frac{(-1)^j}{2j} e^{2i\kappa j} \right].
\end{align}
Since the ordering of the sum is not determined by the physical setup, the limit $N\to\infty$ can produce an arbitrary result. This implies that for any finite $N$ the boundary matters and there is no translation invariance. 

Nevertheless, the chain exhibits directional transport for finite $N$, as can be seen by transforming Eqs.~\eqref{eq:langevin2} into frequency space,
\begin{align}\label{eq:fourierlangevin}
    \sum_{j'=1}^N\left[(\omega_0^2 - \omega^2 - i\gamma_{\rm g}\omega)\delta_{jj'} - \frac{C_{jj'}}{m} \right] z_{j'}[\omega] = \frac{\xi_j[\omega]}{m}.
\end{align}
The directionality of signal transmission is quantified by the mechanical susceptibility of the particle chain, which can be identified from Eq.~\eqref{eq:fourierlangevin} as
\begin{align}\label{eq:susceptibility}
    \big[\chi^{-1}[\omega]\big]_{jj'} = \frac{i}{\omega_0\gamma_{\rm g}}\left[ (\omega^2 - \omega_0^2 - i\gamma_{\rm g}\omega)\delta_{jj'} - \frac{C_{jj'}}{m}\right].
\end{align}
It is normalized such that for $N=1$ particle we have $\chi_{11}[\omega_0]=1$. Directional amplification is present if the susceptibility $\chi_{N1}[\omega]$ for signal transmission from particle $N$ to particle $1$ is much greater than the opposite direction $\chi_{1N}[\omega]$. With the chosen parameters, the particle chain indeed exhibits directional behavior, as shown in Fig.~\ref{fig:sketchmulti} (b). In comparison with the response function of a single particle, resonant signals are significantly enhanced in one direction, while they are always suppressed into the opposite. Note, however, that the particle chain also enhances the recoil noise. A straightforward calculation of the net noise on the $N$-th particle shows that the signal-to-recoil-noise ratio gets worse with increasing particle number. Therefore, the particle chain may enable the detection of very weak signals by overcoming detection noise at the cost of decreasing the signal-to-recoil-noise ratio.

\section{Outlook}\label{conclusion}

We conclude this article by outlining some possible generalizations of the theory presented. 

{\it Rotational optical binding.--} The master equation \eqref{master} describes light-induced torques between co-levitated nanoparticles due to their anisotropic susceptibility tensors. While the results of Sec.~\ref{deeplytrapped} are readily adapted to the libration regime of strongly aligned particles, they must be generalized whenever the inherent nonlinearity of rotations starts to contribute strongly \cite{stickler2021}. Answering questions such as if rotational entanglement can be realized through optical binding requires understanding the quantum dynamics of co-rotating particles in presence of reciprocal and nonreciprocal interactions. The analysis will have to account for both the full nonlinearity of rotations as well as for the possible coupling between the particle rotation and center-of-mass motion.

{\it Nonlinear optical binding.--} Our discussion of quantum optical binding in Sec.~\ref{deeplytrapped} assumes that the optical binding interaction may be linearized in the particle positions. In the absence of strong cooling this approximation will fail eventually if a generalized ${\cal PT}$ symmetry is broken \cite{prl}, since the particle amplitudes increase exponentially with time and nonlinearities of both the trapping potential and the optical-binding interaction become relevant \cite{prl,reisenbauer2023non,livska2023observations}. The two-particle dynamics may then exhibit multistability and stable limit cycles, with great potential for sensing \cite{strogatz2018}. The properties of these limit cycles in the quantum regime and their relation to continuous time crystals \cite{kongkhambut2022} are open questions.

{\it Near-field optical binding.--} A second core assumption in Sec.~\ref{deeplytrapped} is that the distance between the particles is sufficiently large, so that only the far-field contribution to light scattering is relevant. While this is well justified in state-of-the-art setups, near-field optical binding may well become relevant in future experiments if the distances between particles become comparable to the laser wavelength. Investigating light-induced coupling in the near-field requires taking into account that the laser fields levitating two particles at close distances overlap significantly. Whether the tunability of optical binding can be retained in such a situation is still unclear, as is whether the impossibility of generating entanglement via optical binding \cite{prl} carries over to the near field. Naively evaluating the near-field coupling strength would suggest that entanglement can indeed be generated, but this argument fails for commonly used nanoparticles because approximating the scattered fields as homogeneous across neighboring particles cannot be justified at close distances.

{\it Large particles.--} Our theory assumes  the size of the dielectric particles to be small in comparison to the wavelengths of the incoming light fields and the distance between neighboring particles. This approximation allows for deriving the Lagrangian in Sec.~\ref{lightmatterinteraction} which yields the correct conservative light-matter interaction. Generalizing this treatment to situations where the wavelength becomes comparable or even greater than the particle size, where internal Mie resonances \cite{maurer2023,maurer2023quantum} and multiple scattering events become relevant \cite{karasek2008}, is a prerequisite to study quantum optical binding between large objects.

{\it Optical binding in microcavities.--} Microcavities yield strong coupling to nanoparticles \cite{wachter2019}  due to their small mode volume, rendering them attractive for interfacing levitated nanoparticles via optical binding. Adapting the optical binding master equation \eqref{master} to this situation requires the correct electromagnetic field modes (or, equivalently, the Green tensor) for such a highly confined geometry. This will change the master equation in two ways: (i) the  conservative interaction is determined by the adapted Green tensor; (ii) when tracing out the vacuum field, the proper modes must be used rather than the free-space ones to avoid mode overcounting, yielding modified Lindblad operators. 
    
In summary, we derived the quantum theory of light-induced interactions in arrays of levitated nanoparticles, and used it to discuss unidirectional transport in linear particle chains. The framework was also used in Ref. \cite{prl} to identify unique signatures of quantum optical binding, which may be observed in state-of-the-art experiments. We expect that the ability to continuously tune the interaction from fully reciprocal to fully nonreciprocal will render nanoparticle arrays an ideal platform for exploring and exploiting non-Hermitian quantum physics.

\begin{acknowledgements}
H. R., K. H. and B. A. S. acknowledge funding by the Deutsche Forschungsgemeinschaft (DFG, German Research Foundation)--439339706. B. A. S. acknowledges funding by the DFG--510794108 as well as by the Carl-Zeiss-Foundation through the project QPhoton. U. D. acknowledges support from the Austrian Science Fund (FWF, Project DOI 10.55776/I5111).
\end{acknowledgements}

\appendix

\begin{widetext}
\section{Lorentz force and torque on a polarized object}\label{appforce}
This section derives the total force and torque acting on a particle polarized with internal polarization field $\mathbf{P}(\mathbf{r})$ in presence of the electric field $\mathbf{E}(\mathbf{r})$ and magnetic field $\mathbf{B}(\mathbf{r})$, as used in Eq.~\eqref{forcesandtorques}. The associated polarization charge and current densities are $\rho_{\rm P}(\mathbf{r})= -\nabla\cdot\mathbf{P}(\mathbf{r})$ and $\mathbf{j}_{\rm P}(\mathbf{r})=\partial_t \mathbf{P}(\mathbf{r})$, respectively.

The total force $\mathbf{F}$ acting on the particle is obtained by integrating the Lorentz force density \cite{barnett2006,novotny2012}
\begin{align}
    \mathbf{f}_{\rm L}(\mathbf{r}) = -[\nabla\cdot\mathbf{P}(\mathbf{r})]\mathbf{E}(\mathbf{r}) + \partial_t \mathbf{P}(\mathbf{r})\times\mathbf{B}(\mathbf{r})
\end{align}
over the particle volume $\mathcal{V}$. Integration by parts yields
\begin{align}\label{forcedensitypre}
    \mathbf{F} = \int_{\mathcal{V}} d^3\mathbf{r} \Big[ [\mathbf{P}(\mathbf{r})\cdot\nabla]\mathbf{E}(\mathbf{r}) + \partial_t \mathbf{P}(\mathbf{r})\times \mathbf{B}(\mathbf{r}) \Big],
\end{align}
which can be rewritten as
\begin{align}\label{forcedensity}
    \mathbf{F} = \int_{\mathcal{V}} d^3\mathbf{r} \Big[ \nabla' [\mathbf{P}(\mathbf{r})\cdot\mathbf{E}(\mathbf{r}')]_{\mathbf{r}' = \mathbf{r}} + \partial_t [\mathbf{P}(\mathbf{r})\times \mathbf{B}(\mathbf{r})] \Big],
\end{align}
where we used Faraday's law $\nabla\times\mathbf{E}(\mathbf{r}) = -\partial_t \mathbf{B}(\mathbf{r})$ and $\nabla[\mathbf{a}\cdot\mathbf{E}(\mathbf{r})]=(\mathbf{a}\cdot\nabla)\mathbf{E}(\mathbf{r}) + \mathbf{a}\times[\nabla\times\mathbf{E}(\mathbf{r})]$ for a constant vector $\mathbf{a}$. For rapidly oscillating fields, such as for dielectric particles in optical fields, the time derivative averages to zero so that only Eq.~\eqref{force} remains \cite{barnett2006,novotny2012}.

The total torque $\mathbf{N}$ can be obtained by integrating the Lorentz torque density $\mathbf{r}\times\mathbf{f}_{\rm L}(\mathbf{r})$ over the particle volume. Partial integration shows that the total torque \cite{barnett2006}
\begin{align}
    \mathbf{N} = \int_{\mathcal{V}} d^3\mathbf{r} \bigg[ \mathbf{P}(\mathbf{r})\times\mathbf{E}(\mathbf{r}) + \mathbf{r}\times \Big[ [\mathbf{P}(\mathbf{r})\cdot\nabla]\mathbf{E}(\mathbf{r}) + \partial_t \mathbf{P}(\mathbf{r})\times \mathbf{B}(\mathbf{r}) \Big] \bigg],
\end{align}
contains two contributions: (i) The first term describes the intrinsic torque on each volume element; (ii) the second term is the orbital torque density resulting from the effective local force density in Eq.~\eqref{forcedensitypre}. Using Faraday's law, the above vector identity, and averaging the time derivative to zero, one obtains
\begin{align}
    \mathbf{N} = \int_{\mathcal{V}} d^3\mathbf{r} \bigg[ \mathbf{P}(\mathbf{r})\times\mathbf{E}(\mathbf{r}) + \mathbf{r}\times \Big[ \nabla' [\mathbf{P}(\mathbf{r})\cdot\mathbf{E}(\mathbf{r}')]_{\mathbf{r}' = \mathbf{r}}\Big] \bigg].
\end{align}
Subtracting the orbital torque $\mathbf{r}_{\rm cm} \times \mathbf{F}$ acting on the particle center of mass $\mathbf{r}_{\rm cm}$ yields the torque (\ref{torque}).

\section{Approximating the total Lagrange function}\label{appgauge}

This appendix describes the approximations that lead from Eq.~\eqref{lagrangestart} to Eq.~\eqref{Leff}. Explicitly, Eqs.~(\ref{unperturbed}), (\ref{perturbed}) and (\ref{fieldsplit}) are inserted into Eq.~\eqref{lagrangestart}. Using that the external field $\mathbf{A}_{\rm ext}(\mathbf{r},t)$ varies little over the volume of each particle yields
\begin{align}\label{appLtot}
    L_{\rm tot} \approx & L_{\rm m} + L_{\rm em} - V_{\rm ext} + \varepsilon_0\sum_{j=1}^N \int_{\mathcal{V}_j} d^3\mathbf{r}\left( \frac 1 2 \partial_t \mathbf{A}(\mathbf{r})\cdot\upchi_j \partial_t\mathbf{A}(\mathbf{r}) + \partial_t\mathbf{A}(\mathbf{r})\cdot\upchi_j \partial_t \mathbf{A}_{\rm ext}(\mathbf{r},t) \right)\nonumber \\ &+ \varepsilon_0 \sum_{\substack{j,j'=1\\ j\neq j'}}^N \int_{\mathcal{V}_j}d^3\mathbf{r}\int_{\mathcal{V}_{j'}} d^3\mathbf{r}' \left( \frac 1 2  \partial_t \mathbf{A}(\mathbf{r})\cdot\upchi_j {\sf G}_0(\mathbf{r}-\mathbf{r}')\upchi_{j'}\partial_t \mathbf{A}(\mathbf{r}') + \partial_t \mathbf{A}(\mathbf{r})\cdot\upchi_j {\sf G}_0(\mathbf{r}-\mathbf{r}')\upchi_{j'}\partial_t \mathbf{A}_{\rm ext}(\mathbf{r}',t) \right) \\ &+ \int d^3 \mathbf{r}\left( \frac{\varepsilon_0}{2}[\partial_t \mathbf{A}_{\rm ext}(\mathbf{r},t)]^2 - \frac{1}{2\mu_0}[\nabla\times\mathbf{A}_{\rm ext}(\mathbf{r},t)]^2 +\varepsilon_0 \partial_t \mathbf{A}_{\rm ext}(\mathbf{r},t)\cdot\partial_t \mathbf{A}(\mathbf{r}) - \frac{1}{\mu_0} [\nabla\times\mathbf{A}_{\rm ext}(\mathbf{r},t)]\cdot[\nabla\times\mathbf{A}(\mathbf{r})]  \right) \nonumber
\end{align}
Here, $V_{\rm ext}$ collects those terms in which the external field $\mathbf{A}_{\rm ext}(\mathbf{r})$ is integrated over the particle volume only.

We can neglect the light-matter interaction terms that are quadratic in the scattering fields $\mathbf{A}(\mathbf{r})$ since the latter are a small perturbation to the external laser field. In addition, the second term in the second line is negligible when compared to the last term of the first line, which both describe interaction between the external field and the scattering field, since the susceptibility is proportional to the particle volume. The last line of Eq.~\eqref{appLtot} can be simplified by performing a partial integration shifting the curls to $\mathbf{A}_{\rm ext}(\mathbf{r},t)$, which is transverse and fulfills the homogeneous wave equation \eqref{freewave}. Altogether,
\begin{align}\label{Ltotapp}
    L_{\rm tot} \approx L_{\rm m} + L_{\rm em} - V_{\rm ext} - \varepsilon_0 \sum_{j=1}^N \int_{\mathcal{V}_j}d^3\mathbf{r}\Big[ \mathbf{E}_{\rm ext}(\mathbf{r},t)\cdot \upchi_j \partial_t\mathbf{A}(\mathbf{r})\Big] - \varepsilon_0 
\frac{d}{dt}\int d^3\mathbf{r}\left[ \left( \frac 1 2 \mathbf{A}_{\rm ext}(\mathbf{r},t) + \mathbf{A}(\mathbf{r})\right)\cdot \mathbf{E}_{\rm ext}(\mathbf{r},t) \right].
\end{align}
The last term can be gauged away, such that by defining the mechanical gauge function \cite{craig1998}
\begin{align}\label{gaugefunction}
    S = \varepsilon_0 \sum_{j=1}^N \int_{\mathcal{V}_j}d^3\mathbf{r}\left[ \mathbf{E}_{\rm ext}(\mathbf{r},t)\cdot\upchi_j\mathbf{A}(\mathbf{r})\right] + \varepsilon_0 \int d^3\mathbf{r}\left[ \left( \frac 1 2 \mathbf{A}_{\rm ext}(\mathbf{r},t) + \mathbf{A}(\mathbf{r})\right)\cdot \mathbf{E}_{\rm ext}(\mathbf{r},t) \right],
\end{align}
the Lagrangian takes the form
\begin{align}
    L_{\rm tot} = L_{\rm m} + L_{\rm em} - V_{\rm ext} - V_{\rm int} - \frac{dS}{dt} + \varepsilon_0 \sum_{j=1}^N \sum_{q_j} \dot q_j \frac{\partial}{\partial q_j} \int_{\mathcal{V}_j}d^3\mathbf{r}\left[ \mathbf{E}_{\rm ext}(\mathbf{r},t)\cdot\upchi_j\mathbf{A}(\mathbf{r})\right].
\end{align}
If the particles move much slower than the external light field changes, the last term can be neglected and one arrives at $L_{\rm tot}\approx L - dS/dt$ as used in the main text.

\section{Quantum Langevin equations}\label{applangevin}
This Appendix derives the optical binding quantum Langevin equations which are equivalent to the master equation \eqref{master}. They are required to obtain the linearized Langevin equations \eqref{langevinlinear}.

Our starting point is the Heisenberg equations of motion resulting from the total light-matter Hamiltonian \eqref{hamiltonian},
\begin{subequations}\label{heisenbergmomentum}
\begin{align}
    \dot q_j = & \frac{\partial  }{\partial p_j}H_{\rm m}\\
    \dot p_j = & -\frac{\partial}{\partial q_j} (H_{\rm m} + V_{\rm ext} + V_{\rm int}).
\end{align}
\end{subequations}
They depend on the optical degrees of freedom through the interaction potential $V_{\rm int}$. The Heisenberg equations for the light fields,
\begin{align}
    \dot b_{\mathbf{k}s} = -i\omega_k b_{\mathbf{k}s} + \frac{\varepsilon_0 \omega_{\rm L}}{\sqrt{8\hbar\omega_k \varepsilon_0 L^3}} \sum_{j=1}^N \int_{\mathcal{V}_j} d^3\mathbf{r} e^{-i\mathbf{k}\cdot\mathbf{r}} \mathbf{t}_{\mathbf{k}s}^*\cdot\upchi_j[\mathbf{E}_{\rm L}(\mathbf{r})e^{-i\omega_{\rm L} t} - \text{c.c.}],
\end{align}
can be solved as function of the coordinate operators,
\begin{align}
    b_{\mathbf{k}s}(t) = b_{\mathbf{k}s}(t_0) e^{-i\omega_k (t-t_0)} + \frac{\varepsilon_0 \omega_{\rm L}}{\sqrt{8\hbar\omega_k \varepsilon_0 L^3}} \sum_{j=1}^N \int_{t_0}^t dt' \int_{\mathcal{V}_j(t')} d^3\mathbf{r} e^{-i\mathbf{k}\cdot\mathbf{r}} \mathbf{t}_{\mathbf{k}s}^*\cdot\upchi_j(t')[\mathbf{E}_{\rm L}(\mathbf{r})e^{-i\omega_{\rm L} t'} - \text{c.c.}].
\end{align}
Inserting this into Eq.~\eqref{heisenbergmomentum} yields a closed system of Heisenberg equations for the mechanical degrees of freedom.

We now divide the time axis into intervals of width $\Delta t$ much greater than a single optical period $1/\omega_{\rm L}$, but much smaller than the timescale of mechanical motion. Integrating the momentum equations of motion over one time step, theer momentum change  $\Delta p_j(t) = p_j(t+\Delta t) - p_j(t)$ is given by
\begin{align}\label{eqwithnoname}
   \Delta p_j(t) \approx & -\frac{\partial}{\partial q_j} (H_{\rm m} + \bar V_{\rm ext})\Delta t + \xi_q^j \Delta t\nonumber\\ &+ \Bigg[\frac{\partial}{\partial q_j} \int_t^{t+\Delta t}dt'\int_{t}^{t'}dt'' \sum_{\mathbf{k}s}\int_{\mathcal{V}_j(t')}d^3\mathbf{r}\int_{\mathcal{V}_{j'}(t'')}d^3\mathbf{r}'\frac{\omega_{\rm L}^2\varepsilon_0}{8i\omega_k L^3}\nonumber\\ &\times\Big[ e^{i\mathbf{k}\cdot(\mathbf{r}-\mathbf{r}')} e^{-i\omega_k (t'-t'')} \mathbf{t}_{\mathbf{k}s}\cdot\upchi_j(t')[\mathbf{E}_{\rm L}(\mathbf{r})e^{-i\omega_{\rm L} t'} - \text{c.c.}] \mathbf{t}_{\mathbf{k}s}^*\cdot\upchi_{j'}(t'')[\mathbf{E}_{\rm L}(\mathbf{r'})e^{-i\omega_{\rm L} t''} - \text{c.c.}] - \text{H.c.} \Big] \Bigg]_{j'=j}\nonumber\\
   &+ \frac{\partial}{\partial q_j}\sum_{\substack{j'=1\\ j'\neq j}}^N \int_t^{t+\Delta t}dt'\int_{t}^{t'}dt'' \sum_{\mathbf{k}s}\int_{\mathcal{V}_j(t')}d^3\mathbf{r}\int_{\mathcal{V}_{j'}(t'')}d^3\mathbf{r}'\frac{\omega_{\rm L}^2\varepsilon_0}{8i\omega_k L^3}\nonumber\\
   &\times \Big[ e^{i\mathbf{k}\cdot(\mathbf{r}-\mathbf{r}')} e^{-i\omega_k (t'-t'')} \mathbf{t}_{\mathbf{k}s}\cdot\upchi_j(t')[\mathbf{E}_{\rm L}(\mathbf{r})e^{-i\omega_{\rm L} t'} - \text{c.c.}] \mathbf{t}_{\mathbf{k}s}^*\cdot\upchi_{j'}(t'')[\mathbf{E}_{\rm L}(\mathbf{r'})e^{-i\omega_{\rm L} t''} - \text{c.c.}] - \text{H.c.} \Big],
\end{align}
where $\bar V_{\rm ext}$ is the time-averaged external potential $V_{\rm ext}$. The radiation pressure shot noise operators
\begin{align}\label{xi}
    \xi_q^j\Delta t = \frac{\partial}{\partial q_j}\int_t^{t+\Delta t}dt' \sum_{\mathbf{k}s}\int_{\mathcal{V}_j(t')}d^3\mathbf{r} \frac{\omega_{\rm L}}{2i}\sqrt{\frac{\hbar\varepsilon_0}{2\omega_k L^3}} \left[ \mathbf{t}_{\mathbf{k}s}e^{i\mathbf{k}\cdot\mathbf{r}} e^{-i\omega_k (t'-t)} b_{\mathbf{k}s}(t) + \text{H.c.} \right]\cdot\upchi_j(t')[\mathbf{E}_{\rm L}(\mathbf{r})e^{-i\omega_{\rm L} t'} - \text{c.c.}]
\end{align}
describe the generalized forces due to the light-matter interaction when ignoring the backaction of the scattered light on the particles. The term in the second and third line describes the effect of radiation pressure on the $j$th particle, while the fourth and fifth line describes the scattering contribution to the optical binding interaction between particles $j$ and $j'$.

We now perform the Markov approximation by replacing all coordinate operators as $q_j(t') \simeq q_j(t)$. Then we evaluate all integrals over $t'$ and $t''$ by using the relations in Sec.~\ref{traceout}, such as $\pi\delta(\omega_{\rm L}-\omega_k) + i\mathcal{P}[1/(\omega_{\rm L}-\omega_k)] = 1/i(\omega_k - \omega_{\rm L} - i\eta)$. Utilizing Eq.~\eqref{residue}, one can rewrite the following integral as
\begin{align}
    &\int_{\mathcal{V}_j}d^3\mathbf{r}\int_{\mathcal{V}_{j'}}d^3\mathbf{r}' \int d^3\mathbf{k}\sum_s\frac{k_{\rm L}^2 \varepsilon_0\Delta t}{4 (2\pi)^3}\left[ e^{i\mathbf{k}\cdot(\mathbf{r}-\mathbf{r}')} \mathbf{t}_{\mathbf{k}s}\cdot\upchi_j\mathbf{E}_{\rm L}^*(\mathbf{r}) \frac{1}{k^2-k_{\rm L}^2-i\eta} \mathbf{t}_{\mathbf{k}s}^*\cdot\upchi_{j'}\mathbf{E}_{\rm L}(\mathbf{r}') +\text{H.c.}\right]\nonumber\\ &
    = \int_{\mathcal{V}_j}d^3\mathbf{r}\int_{\mathcal{V}_{j'}}d^3\mathbf{r}' \frac{\varepsilon_0\Delta t}{4} \Big[ \mathbf{E}_{\rm L}^*(\mathbf{r})\cdot\upchi_j [{\sf G}(\mathbf{r}-\mathbf{r'})-{\sf G}_0(\mathbf{r}-\mathbf{r'})]\upchi_{j'}\mathbf{E}_{\rm L}(\mathbf{r}) +\text{H.c.}\Big].
\end{align}
For $j'\neq j$ the volume integrals can be replaced by the respective total volume, and for $j' = j$ Eq.~\eqref{approxGreen} holds.

The shot noise force operators \eqref{xi} are not necessarily Gaussian. Its first and second moments are
\begin{subequations}
\begin{align}
    \langle \xi_q^j(t) A_{\rm mec} \rangle & = 0\\
    \langle \xi_{q'}^{j'}(t') A_{\rm mec} \xi_q^j(t) B_{\rm mec} \rangle & = \int d^3\mathbf{k}\sum_s \frac{\hbar\varepsilon_0 \omega_{\rm L}^2}{8\omega_k (2\pi)^3} \text{sinc}^2\left(\frac{\omega_{\rm L}-\omega_k}{2}\Delta t\right) e^{-i(\omega_k - \omega_{\rm L})(t'-t)} \label{firstmoment}\\ &\times\Bigg\langle\left[ \frac{\partial}{\partial q'_{j'}}\int_{\mathcal{V}_{j'}}d^3\mathbf{r}' \, \mathbf{t}_{\mathbf{k}s}\cdot\upchi_{j'}\mathbf{E}_{\rm L}^*(\mathbf{r}')e^{i\mathbf{k}\cdot\mathbf{r}'} \right] A_{\rm mec}  \left[ \frac{\partial}{\partial q_{j}}\int_{\mathcal{V}_j}d^3\mathbf{r} \, \mathbf{t}_{\mathbf{k}s}^*\cdot\upchi_j\mathbf{E}_{\rm L}(\mathbf{r})e^{-i\mathbf{k}\cdot\mathbf{r}} \right] B_{\rm mec}\Bigg\rangle,\nonumber
\end{align}
\end{subequations}
for arbitrary operator $A_{\rm mec}$ and $B_{\rm mec}$ acting in the mechanical Hilbert space. Here, we took the electromagnetic field to be in the vacuum at $t$. 

We now use that for functions $f(\omega_{k})$ varying slowly in relation to $1/\Delta t$,
\begin{align}
    \int_{0}^\infty d\omega_k\, f(\omega_k) \text{sinc}^2\left( \frac{\omega_{\rm L}-\omega_k}{2}\Delta t \right) e^{-i(\omega_k-\omega_{\rm L})\tau} \approx  f(\omega_{\rm L})\int_{0}^\infty d\omega_k\, \text{sinc}^2\left( \frac{\omega_{\rm L}-\omega_k}{2}\Delta t \right) e^{-i(\omega_k-\omega_{\rm L})\tau} \approx 2\pi f(\omega_{\rm L})\delta(\tau),
\end{align}
to get
\begin{align}\label{secondmoment}
    \langle \xi_{q'}^{j'}(t')A_{\rm mec} \xi_q^j(t) B_{\rm mec} \rangle =& \int d^2\mathbf{n}\sum_s \frac{\hbar\varepsilon_0 k_{\rm L}^3}{32\pi^2}\delta(t-t')\nonumber\\ &\times\left\langle \left[ \frac{\partial}{\partial q'_{j'}} V_{j'} \mathbf{t}_{\mathbf{n}s}\cdot\upchi_{j'}\mathbf{E}_{\rm L}^*(\mathbf{r}_{j'}) e^{ik_{\rm L}\mathbf{n}\cdot\mathbf{r}_{j'}} \right]A_{\rm mec} \left[ \frac{\partial}{\partial q_{j}} V_{j} \mathbf{t}_{\mathbf{n}s}^*\cdot\upchi_{j}\mathbf{E}_{\rm L}(\mathbf{r}_{j}) e^{-ik_{\rm L}\mathbf{n}\cdot\mathbf{r}_{j}}\right]B_{\rm mec}\right\rangle.
\end{align}
The commutators of the noise force operators can be obtained in a similar manner. In the limit of small time steps, Eqs.~\eqref{eqwithnoname} turn into the quantum Langevin equations of optical binding,
\begin{subequations}\label{langevinfull}
\begin{align}
    \dot q_j =& \frac{\partial}{\partial p_j} H_{\rm m} \\
    \dot p_j =& -\frac{\partial}{\partial q_j} (H_{\rm m} + V_{\rm L})  + \frac{\varepsilon_0 k_{\rm L}^3 V_j}{12\pi}\text{Im}\left[ \mathbf{E}_{\rm L}^*(\mathbf{r}_j)\cdot\upchi_j \frac{\partial}{\partial q_j} V_j \upchi_j \mathbf{E}_{\rm L}(\mathbf{r}_j)\right] \nonumber\\ &+ \frac{\partial}{\partial q_j} \sum_{\substack{j'=1\\ j'\neq j}}^N \frac{\varepsilon_0 V_j V_{j'}}{2} \text{Re}\left[ \mathbf{E}_{\rm L}^*(\mathbf{r}_j)\cdot\upchi_j {\sf G}(\mathbf{r}_j-\mathbf{r}_{j'})\upchi_{j'}\mathbf{E}_{\rm L}(\mathbf{r}_{j'}) \right] + \xi_q^j.
\end{align}
\end{subequations}
The expectation value of these equations yields the averaged classical optical binding equations of motion, whose center-of-mass version was derived in \cite{dholakia2010,rieser2022}. The same equations are obtained from the Ehrenfest equations resulting from the optical binding master equation \eqref{master}. This confirms the equivalence between the optical binding master equation \eqref{master} and the quantum Langevin equations \eqref{langevinfull}.

The noise operators in \eqref{langevinfull} are characterized by their first and second moments \eqref{firstmoment} and \eqref{secondmoment} as well as all higher moments following from the definition \eqref{xi}. In the regime of linear harmonic motion, the first two moments suffice to characterize the noise, which is thus Gaussian. In order to calculate its correlator, one requires the second moments with $A_{\rm mec}=\mathbb{1}$. In this case, the integral over all scattering directions $\mathbf{n}$ can be evaluated explicitly, so that for $j \neq j'$,
\begin{align}\label{correlationsfull1}
    \langle \xi_{q'}^{j'}(t')\xi_q^j(t) B_{\rm mec} \rangle = \frac{\hbar\varepsilon_0}{8} \delta(t-t') \left\langle\frac{\partial^2}{\partial q'_{j'} \partial q_j} V_j V_{j'} \mathbf{E}_{\rm L}^*(\mathbf{r}_{j'})\cdot\upchi_{j'}\text{Im}[{\sf G}(\mathbf{r}_j-\mathbf{r}_{j'})]\upchi_j\mathbf{E}_{\rm L}(\mathbf{r}_j)B_{\rm mec}\right\rangle,
\end{align}
while for $j=j'$,
\begin{align}\label{correlationsfull}
    \langle \xi_{q'}^j(t')\xi_q^j(t) B_{\rm mec} \rangle =& \frac{\hbar\varepsilon_0 k_{\rm L}^3}{60\pi}\delta(t-t')\Bigg\langle \Bigg[ 5 \bigg[\frac{\partial}{\partial q'_{j}} V_{j} \upchi_{j}\mathbf{E}_{\rm L}^*(\mathbf{r}_{j}) \bigg]\cdot\bigg[\frac{\partial}{\partial q_{j}} V_{j} \upchi_{j}\mathbf{E}_{\rm L}(\mathbf{r}_{j})\bigg]+ 2 k_{\rm L}^2 \frac{\partial \mathbf{r}_j}{\partial q'_{j}}\cdot \frac{\partial \mathbf{r}_j}{\partial q_{j}} | V_{j} \upchi_{j}\mathbf{E}_{\rm L}(\mathbf{r}_{j}) |^2\nonumber\\&  - k_{\rm L}^2 \bigg[ V_{j} \frac{\partial \mathbf{r}_j}{\partial q'_{j}} \cdot\upchi_{j}\mathbf{E}_{\rm L}^* (\mathbf{r}_{j}) \bigg] \bigg[ V_{j} \frac{\partial \mathbf{r}_j}{\partial q_{j}} \cdot\upchi_{j}\mathbf{E}_{\rm L}(\mathbf{r}_{j}) \bigg] \Bigg] B_{\rm mec} \Bigg\rangle.
\end{align}
For $B_{\rm mec}=\mathbb{1}$ and for $q = q'$ given by the equilibrium coordinates in an optical tweezer, this last expression reduces to the recoil diffusion rates of small ellipsoidal rotors \cite{schafer2021}.

To get the linearized Langevin equations \eqref{langevinlinear} between small spheres, Eqs.~\eqref{langevinfull} must be expanded harmonically around the tweezer foci following the same steps as in Sec.~\ref{sec:array} and renaming $\xi_j = \xi_{z}^j$. The corresponding correlators \eqref{correlator} are obtained by evaluating Eqs.~\eqref{correlationsfull1} and \eqref{correlationsfull} at the tweezer foci.

\end{widetext}

\bibliographystyle{myapsrev}

\end{document}